\documentclass[a4paper,11pt]{article}
\usepackage{jheppub} 
\usepackage{lineno}
\usepackage{comment}
\usepackage{graphicx}
\usepackage{subcaption}
\usepackage{xcolor} 

\title{\boldmath A Thermodynamic Positivity Bound on Higher-Derivative 3-Form Couplings in de Sitter, and its Inflationary Consequences}

\author[a]{Nutthaphat Lunrasri, }
\author[a]{Chakrit Pongkitivanichkul}
\affiliation[a]{Khon Kaen Particle Physics and Cosmology Theory Group (KKPaCT),\\ Department of Physics, Faculty of Science, Khon Kaen University, 123 Mitraphap Rd.,\\ Khon Kaen, 40002, Thailand}

\emailAdd{natthapatl@kkumail.com}
\emailAdd{chakpo@kku.ac.th}

\abstract{
We investigate the interplay between the thermodynamic positivity bounds
and slow-roll inflation within a framework governed by a 3-form gauge
field. Starting from classical considerations, we derive an upper bound on
the mass of black holes in dS spacetime which constrains the admissible
parameter space. To incorporate quantum gravity effects, we introduce
higher-derivative corrections to the 3-form action and, by requiring the
Wald entropy correction to be positive, obtain a strict bound on these
terms. Evaluating the backreaction within a quasi-local thermodynamic
cavity bounded by the zero-force surface, we find that the correction to
the extremal mass vanishes, so that the exact Nariai state saturates the
classical bound rather than being shifted below it. The resulting bound is
found to be invariant under field redefinitions of the metric. Extending
this setup to
cosmological inflation, we examine the scalar dual of the 3-form in both
large-field and small-field regimes. In the large-field limit, the
potential acquires a Higgs-like structure that supports slow-roll inflation
consistent with Planck data. In contrast, the small-field limit leads to
an effective potential with an AdS minimum, rendering it inconsistent with
the dS swampland constraints. Notably, we find that thermodynamic
consistency can impose constraints more stringent than those derived from
inflationary dynamics alone. These results underscore the utility of
swampland-inspired principles in shaping viable models of early universe
cosmology.
}

\begin{document}
\maketitle
\flushbottom

\section{Introduction}
String/M theory plays a crucial role in the study of quantum gravity. The central idea is that fundamental particles arise as quantized excitations of higher-dimensional objects, known as p-branes, which require extra dimensions to maintain Lorentz symmetry. The low-energy effective action is derived through dimensional reduction, achieved by considering suitable compactifications that lead to various possible vacua. This vast landscape of vacua is explored within the swampland program proposed by \cite{Vafa:2005ui}, which aims to identify criteria that distinguish consistent quantum gravity theories from inconsistent ones.

A cornerstone of this program is the Weak Gravity Conjecture (WGC) \cite{Arkani-Hamed:2006emk}, which dictates that gravity must remain the weakest force, a principle that has also been generalized to de Sitter spacetimes \cite{Antoniadis:2020xso}. While the WGC is traditionally framed around the kinematic decay of extremal charged objects, another rigorous theoretical pathway to constrain effective field theories involves establishing bounds on quantum corrections to black hole solutions. For instance, \cite{Kats:2006xp} studies a charged black hole with higher-order interactions between a 1-form gauge field and gravity, interpreting the resulting quantum corrections as backreaction effects to constrain coupling constants. This approach was crucially refined in \cite{Cheung:2018cwt} using black hole thermodynamics, demonstrating that evaluating the Wald entropy leads to a new formulation where $\Delta S > 0$ provides a strict positivity bound on higher-order couplings. This thermodynamic consistency requirement offers a powerful, independent metric for delineating viable quantum gravity theories from the swampland.

In string/M-theory, $p$-branes naturally couple to $(p+1)$-form gauge fields, generalizing the familiar coupling of point particles to 1-form fields. For instance, in Type IIA string theory, the 3-form field couples to membranes, while in M-theory, fundamental objects such as M2-branes and M5-branes couple to 3-form and 6-form fields, respectively \cite{Ovrut:1997ur,Bandos:2018gjp}. Upon compactification to four dimensions, these higher-form fields give rise to a rich landscape of effective theories, depending on the geometry and fluxes of the internal manifold. Given the multitude of possible compactifications, it becomes essential to establish criteria—such as generalizations of the WGC—to delineate consistent theories from those in the swampland.

Building on this thermodynamic approach, subsequent work sharpened and generalized the entropy-extremality bound. A universal relation was established linking higher-derivative corrections to black hole entropy with corrections to the extremality bound~\cite{Goon:2019faz}, later extended to extremal (zero-temperature) black holes~\cite{McPeak:2021gpt}. These results built on earlier calculations of higher-derivative corrections to Kerr black holes~\cite{Reall:2019sec} and to AdS-Reissner-Nordstr\"om black holes~\cite{Cremonini:2019xue}, while independent proofs of the WGC from positivity and unitarity/causality constraints~\cite{Cheung:2016wjt,Hamada:2018dde} and an application to dilatonic black holes~\cite{Loges:2020trf} further reinforced this thermodynamic perspective. More recently, these bounds have been tested against increasingly exotic black hole backgrounds, including ModMax electrodynamics coupled to $F(R)$ gravity~\cite{NooriGashti:2025modmax}, Rastall gravity with a surrounding perfect fluid~\cite{NooriGashti:2024rastall}, and charged rotating AdS black holes surrounded by quintessence and string clouds~\cite{Sadeghi:2024wgc}, alongside the related de Sitter swampland conjecture in string field inflation~\cite{Sadeghi:2023dstring}. Interestingly, these higher-form gauge fields and their associated branes also play a direct role outside the black hole context: three-form fields, in particular, have been embedded in braneworld scenarios such as a type-II Randall-Sundrum setup~\cite{Barros:2015evi}, motivating a closer look at their cosmological applications.

Beyond quantum gravity, 3-form fields have also been explored in the context of cosmological inflation \cite{Koivisto:2009fb,Germani:2009iq,Mulryne:2012ax,Kaloper:2011jz}. A key feature of 3-form inflation is the existence of stable attractor solutions, which allow inflation to occur over a wide range of initial conditions. This robustness enhances its viability as an inflationary model. Moreover, 3-form-driven inflation can yield distinctive observational signatures, such as modifications to the CMB anisotropies and enhanced non-Gaussianities, potentially distinguishing it from standard scalar field models.

In this work, we propose a swampland criterion based on the dynamics of a 3-form gauge field, derived from thermodynamic positivity bounds. This criterion is motivated by the thermodynamics of black holes in the presence of a 3-form field and a positive cosmological constant. We emphasize at the outset that this setup differs from the charged black holes underlying the WGC literature above: on the black-hole background the 3-form flux is fixed by the equations of motion to the cosmological constant itself, so there is no conserved gauge charge and no charge-to-mass ratio for quantum corrections to shift, and the extremality relation is the Nariai bound rather than a mass-versus-charge inequality. We therefore do not frame our result as a Weak Gravity Conjecture, and our bound follows from entropy positivity alone rather than from any decay argument. From the requirement that such black holes satisfy the standard thermodynamic condition $\Delta S > 0$, we obtain a thermodynamic positivity bound on the higher-derivative couplings, and we investigate whether this relation remains robust under quantum corrections. 

To explore this question, we introduce higher-derivative corrections to the 3-form action and examine their effects in two complementary ways. First, we evaluate the Wald entropy including higher-derivative corrections and extract the corresponding constraints on the higher-order couplings; we show, moreover, that the resulting bound is invariant under field redefinitions of the metric, which we take as the principal evidence that it is physical. Second, we study the shift in the extremality condition. Evaluating the backreaction within a quasi-local thermodynamic cavity bounded by the zero-force surface, we find that the higher-derivative correction to the extremal mass vanishes identically: the cavity volume shrinks to zero in the degenerate-horizon limit, so the exact Nariai state saturates the classical mass bound rather than being shifted below it, and a non-zero shift arises only in the near-extremal regime where the cavity volume remains finite. The bound therefore rests entirely on entropy positivity, not on extremality. 

We then apply our higher-order 3-form framework to cosmological inflation, mapping the 3-form action to an effective scalar potential through the standard 4-form/scalar duality \cite{Aurilia:1980xj,Dvali:2005an,Kaloper:2008fb}, whose relation to our black-hole analysis we discuss in Section~5. In the large-field regime, the potential takes on a Higgs-like form, enabling us to analyze the slow-roll conditions and derive bounds on inflationary parameters consistent with Planck data. In the small-field limit, the potential reduces to a quartic form whose minimum corresponds to an anti-de Sitter (AdS) vacuum. This feature renders the solution incompatible with the dS swampland criterion, which disfavors stable AdS vacua in quantum gravity. Notably, we find that these thermodynamic bounds may yield stronger theoretical constraints than those arising from observational slow-roll criteria alone.

This paper is organized as follows. In Section~2, we review the classical black hole solution supported by a 3-form gauge field. In Section~3, we incorporate higher-order corrections and analyze both the modified geometry and entropy. The energy condition is also verified in this section. Section~4 presents the inflationary dynamics under our 3-form framework. The key results are summarized and discussed in Section~5, where we also discuss future directions and implications for quantum gravity.

\section{3-form Black Hole solution}

We begin with a review of the classical solution of a black hole supported by a 3-form gauge field, following the formalism and results presented in \cite{Barros:2020ghz}. Consider the following action:
\begin{equation}
\label{s0}
    S_0 = \int d^4 x \sqrt{-g}\left( \frac{R}{2 \kappa^2}-\frac{1}{48 {g_{3}}^{2}}F_{\mu\nu\sigma\rho}F^{\mu\nu\sigma\rho}+V(A^2)\right),
\end{equation}
where $V(A^2)$ is the 3-form potential and $g_{3}$ represents a 3-form coupling constant. The first term is the Einstein-Hilbert action, and the second term is the 4-form field strength tensor, expressed in terms of the 3-form gauge field $A_{\nu\sigma\rho}$ field as follows
\begin{equation}
    F_{\mu\nu\sigma\rho} = 4! \nabla_{[\mu}A_{\nu\sigma\rho]}.
\end{equation}
The variation of the action \eqref{s0} with respect to $g_{\mu\nu}$ provides the Einstein field equation,
\begin{align}
    R_{\mu}^{\ \ \nu} = \kappa^2 \left( T_{\mu}^{\ \ \nu} -\frac{1}{2}\delta_{\mu}^{\ \nu}T\right),
\end{align}
where the energy-momentum tensor of the 3-form gauge field $T_{\mu}^{\ \ \nu}$ is written as
\begin{equation}
    \begin{aligned}
     T_{\mu}^{\ \nu} = \frac{1}{6 {g_{3}}^{2}}F_{\mu\sigma\rho\lambda}F^{\nu\sigma\rho\lambda}+6\frac{\partial V}{\partial (A^2)}A_{\mu\sigma\rho}A^{\nu\sigma\rho}+\delta_{\mu}^{\ \nu}\left( -\frac{1}{48 {g_{3}}^{2}}F_{\mu\nu\sigma\rho}F^{\mu\nu\sigma\rho}+V(A^2)\right).
    \end{aligned}
\end{equation}
For a spherically symmetric and static spacetime, the metric is written as
\begin{equation}
    ds^2 = - e^{\alpha(r)}dt^2+e^{\beta(r)}dr^2+r^2d\theta^2+r^2 \sin^2{\theta}d\phi^2,
\end{equation}
which provides geometric parts listed as
\begin{equation}
\label{Ricci}
    \begin{aligned}
        R_{0}^{\ 0} = e^{-\beta (r)} \left(\frac{1}{4} \left(-2 \alpha ''(r)+\alpha '(r)\beta '(r)-\alpha '(r)^{2}\right)-\frac{\alpha '(r)}{r}\right)\\
         R_{1}^{\ 1} = e^{-\beta (r)} \left(\frac{1}{4} \left(-2 \alpha ''(r)+\alpha '(r)\beta '(r)-\alpha '(r)^{2}\right)+\frac{\beta '(r)}{r}\right)\\
         R_{2}^{\ 2}=R_{3}^{\ 3} = \frac{e^{-\beta (r)}}{2 r^2}\left(-r \alpha '(r)+r \beta '(r)+2 e^{\beta (r)}-2\right),
    \end{aligned}
\end{equation}
where the prime symbol is used for the differentiation with respect to $r$.
In the absence of the potential, $V(A^2) = 0$, the relation between $\alpha(r)$ and $\beta(r)$ can be integrated from the combination of Einstein field equations as
\begin{equation}
\label{a-b}
    \alpha(r) = -\beta(r),
\end{equation}
where the constant of integration is set to be zero.

Using the dual field representation, the 3-form can be constructed from the Hodge dual of a 1-form $B^{\mu}$, expressed as
\begin{equation}
    A_{\alpha\beta\gamma} = \sqrt{-g}\epsilon_{\alpha\beta\gamma\lambda}B^{\lambda}(r)
\end{equation}
The components of the dual vector are defined as
\begin{equation}
    B^{\lambda}(r) = \left(0,\xi(r),0,0\right)^{T}.
\end{equation}
Therefore, the non-vanishing components of the 3-form are only $A_{023}$ and the field strength tensor can be written as
\begin{equation}
    F_{\alpha 1 \gamma\delta} = -\partial_{r}A_{\gamma\delta\alpha} = -\partial_{r}\left( \sqrt{-g}\epsilon_{\gamma\delta\alpha\lambda}B^{\lambda}\right)= -\partial_{r}\left( \sqrt{-g}\epsilon_{\gamma\delta\alpha 1}\xi(r)\right),
\end{equation}
which implies that the non-vanishing component of the 4-form tensor is $F_{0123}$, given by
\begin{equation}
    F_{0123} = -\partial_{r}\left(\sqrt{-g}\xi(r)\right) = -\sqrt{-g}\left( \xi'(r)+\frac{1}{2}\left(\alpha'(r)+\beta'(r)+\frac{4}{r}\right)\xi(r)\right).
\end{equation}
The contravariant form is then given by 
\begin{equation}
\label{Fuuuu}
    F^{0123} = \frac{1}{g}F_{0123} = \frac{1}{\sqrt{-g}}\left( \xi'(r)+\frac{1}{2}\left(\alpha'(r)+\beta'(r)+\frac{4}{r}\right)\xi(r)\right).
\end{equation}
The equation of motion for $\xi(r)$ can be obtained by varying the action with respect to the 3-form $A_{\mu\nu\sigma}$, resulting in the following expression
\begin{align}
\label{EOM3}
    \nabla_{\mu}F^{\mu\nu\sigma\rho} = 0.
\end{align}
By plugging equation \eqref{Fuuuu}, the previous equation becomes
\begin{equation}
\label{xieq}
    \xi''(r)+\frac{1}{2}\left(\alpha'(r)+\beta'(r)+\frac{4}{r}\right)\xi'(r)+\frac{1}{2}\left(\alpha''(r)+\beta''(r)-\frac{4}{r^2}\right)\xi(r) =0.
\end{equation}
From \eqref{a-b}, the previous equation can be re-expressed as 
\begin{align}
     \xi''(r)+\frac{2}{r} \xi'(r)-\frac{2}{r^2}\xi(r) =0.
\end{align}
The general solution is written as
\begin{equation}
    \xi(r) = c_{0}r+\frac{\tilde{c}_{0}}{r^2},
\end{equation}
and the kinetic term can be simplified as
\begin{equation}
        \frac{1}{48 {g_{3}}^{2}}F_{\mu\nu\sigma\rho}F^{\mu\nu\sigma\rho} =-\frac{9}{2 {g_{3}}^{2}} c_{0}^2.
\end{equation}

Another combination of the Einstein field equations yields the following.
\begin{equation}
    \alpha'' + \alpha'^2 = \frac{2}{r^2} \left(1 - e^{-\alpha}\right).
\end{equation}
By defining \(\alpha = \ln f\), the previous equation becomes
\begin{equation}
    f'' - \frac{2f}{r^2} + \frac{2}{r^2} = 0.
\end{equation}
The solution of this differential equation is given by
\begin{equation}
\label{f0}
    f(r) = 1 + \frac{K_1}{r} + K_2 r^2,
\end{equation}
where $K_1$ and $K_2$ are integration constants. To determine these constants, we use the relation
\begin{equation}
    R_{0}^{\ 0} = \kappa^2\left( T_{0}^{\ 0}-\frac{1}{2}T   \right),
\end{equation}
and substitute equation \eqref{f0} into the left-hand side of the above equation. This yields the result
\begin{equation}
   - K_{2} =  \frac{3c_{0}^{2}\kappa^{2}}{2 {g_{3}^{2}}}
\end{equation}
which imposes the condition that $K_{2}$ must be negative to ensure that $c_{0}$ is a real number. Equation \eqref{f0} can then be rewritten as
\begin{equation}
    f(r) = 1+\frac{K_{1}}{r} - \frac{3c_{0}^2\kappa^{2}}{2 {g_{3}}^{2}}r^2.
\end{equation}
In the absence of a matter field, where $\frac{1}{48}F^2 \equiv \frac{1}{48} F_{\mu\nu\rho\sigma}F^{\mu\nu\rho\sigma} = 0$, the constant $c_{0}$ must vanish, and the metric should reduce to the Schwarzschild solution. Therefore, we impose
\begin{equation}
    K_{1} =-\frac{\kappa^2 m}{8 \pi} =-{\kappa}^2 M.
\end{equation}
The resulting metric is given by
\begin{equation}
\label{solutionmetric}
    e^{-\beta(r)} = 1-\frac{\kappa^2 M}{r}- \frac{3c_{0}^2\kappa^{2}}{2 {g_{3}}^{2}}r^2,
\end{equation}
which implies that the 3-form gauge field endows the black hole with de Sitter-like properties by imposing 
\begin{equation}
    \frac{\Lambda}{3}=\frac{1}{\ell^{2}} = \frac{3c_{0}^2\kappa^{2}}{2 {g_{3}}^{2}},\label{eq:lambda}
\end{equation}
where $\Lambda$ represents the cosmological constant, and $\ell$ is the de Sitter radius.

The previous solution consists of three event horizons: one complex and two real positive horizons. The real positive ones can be expressed as follows \cite{Fernando:2013mex}
\begin{equation}
\begin{aligned}
      r_H &= \frac{2 \ell}{\sqrt{3}} \cos\left(\frac{\gamma}{3} + \frac{4\pi}{3}\right), \\
r_c &= \frac{2 \ell}{\sqrt{3}} \cos\left(\frac{\gamma}{3}\right),  
\end{aligned}
\end{equation}
where
\begin{equation}
    \gamma = \cos^{-1}\left(- \frac{3\sqrt{3}\kappa^2 M}{2\ell}\right).
\end{equation}
The existence of these horizons is guaranteed by the condition:
\begin{figure}[htbp]
\centering
\includegraphics[scale=0.8]{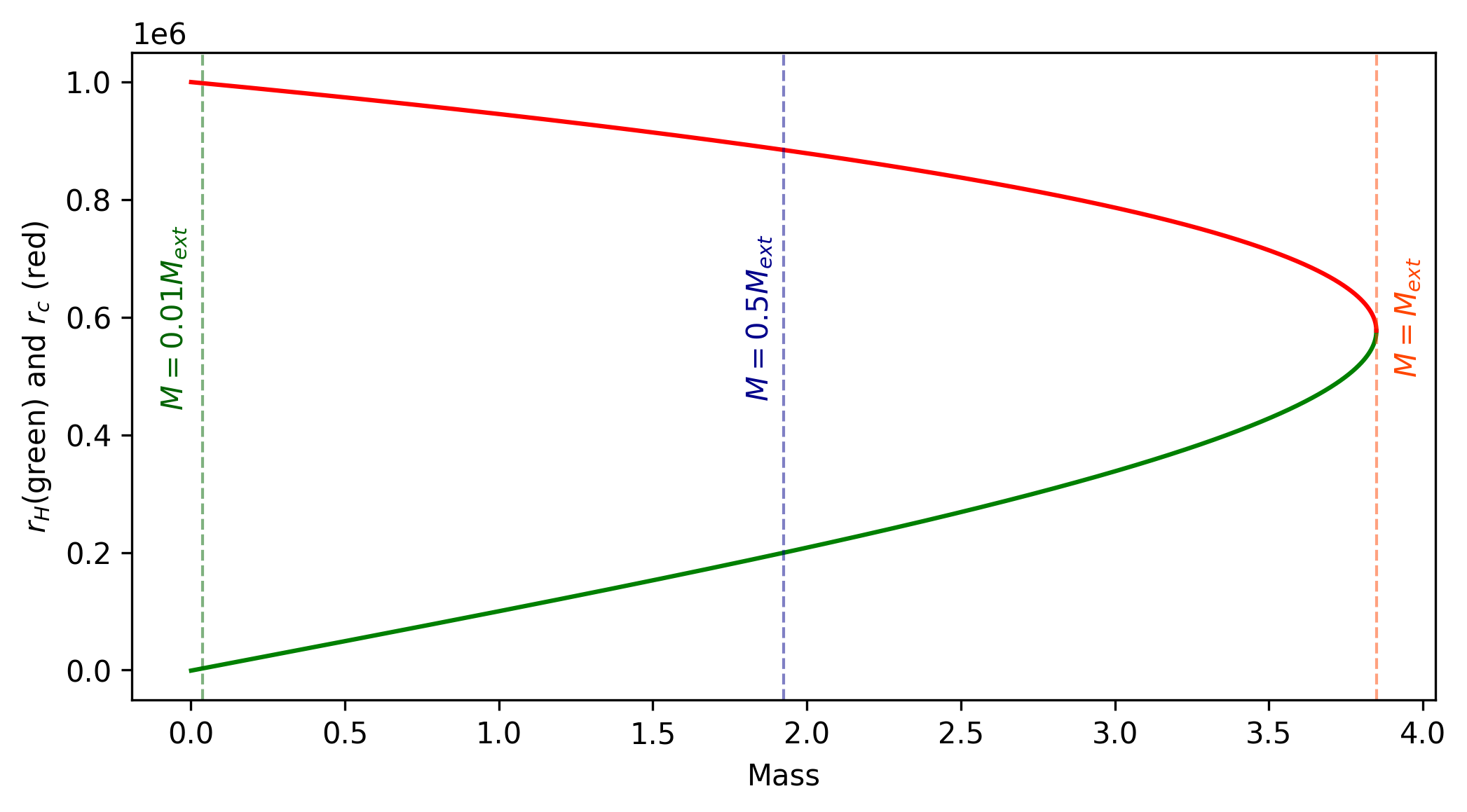}
\caption{ The plot illustrates the dependence of the event horizon radius \( r_H \) (green curve) and the cosmological horizon radius \( r_c \) (red curve) on the black hole mass. 
        The vertical dashed lines represent specific values of the black hole mass expressed as fractions of the extremal mass \( M_{\text{ext}} \): 
        \( M = 0.01\,M_{\text{ext}} \) (green), \( M = 0.5\,M_{\text{ext}} \) (blue), and \( M = M_{\text{ext}} \) (orange). 
        As the mass increases toward \( M_{\text{ext}} \), the event and cosmological horizons approach each other, eventually coinciding at the extremal limit.\label{H1}} 
\end{figure}
\begin{equation}
\label{b1}
    \kappa^4 M^2 \leq \frac{4 \ell^2}{27},
\end{equation}
where the range of radius direction is $0 < r_H \leq r_c$. The relation between $r_H$ and $r_c$ satisfying the above condition can be illustrated in Figure \ref{H1}. The extremal limit corresponds to $\kappa^4 M^2 = \frac{4 \ell^2}{27}$, where the horizon coincides with the cosmological horizon $(r_H = r_c = 3 \kappa^2 M/2)$.

\section{Black Hole solution with correction term}
In this section, we extend the three-form model by introducing higher-order terms to incorporate quantum gravity corrections into the effective action and analyze the resulting entropy.

\subsection{Geometric Solution}
The action \eqref{s0} including correction terms is written as
\begin{equation}
\label{paction}
\begin{aligned}
    S = &\int d^4 x \sqrt{-g}\bigg( \frac{R}{2 \kappa^2}-\frac{1}{48{g_{3}}^{2}}F_{\mu\nu\sigma\rho}F^{\mu\nu\sigma\rho}+c_1 R^2+c_2 R_{\mu\nu}R^{\mu\nu}+c_3 R_{\mu\nu\sigma\rho}R^{\mu\nu\sigma\rho} \\
    &\quad\quad\quad\quad\quad\quad+c_4 RF_{\mu\nu\sigma\rho}F^{\mu\nu\sigma\rho}+c_5 R^{\mu\nu}F_{\mu}^{\ \sigma\rho\delta}F_{\nu\sigma\rho\delta}+c_6 R^{\mu\nu\sigma\rho}F_{\mu\nu}^{\ \ \lambda\delta}F_{\sigma\rho\lambda\delta}  \\
    &\quad\quad\quad\quad\quad\quad+ c_7 (F_{\mu\nu\sigma\rho}F^{\mu\nu\sigma\rho} )^2+c_8 \nabla_{\alpha}F_{\mu\nu\sigma\rho}\nabla^{\alpha}F^{\mu\nu\sigma\rho}\bigg),
\end{aligned}
\end{equation}
where $(F_{\mu\nu\sigma\rho}F^{\mu\nu\sigma\rho} )^2 = F_{\mu\nu\sigma\rho}F^{\mu\nu\sigma\rho} F_{\alpha\beta\lambda\delta}F^{\alpha\beta\lambda\delta}$. Additionally, we treat the quantum gravity correction as the black reaction effect, as will be shown below.

\begin{figure}[htbp]
\centering
\includegraphics[scale=0.6]{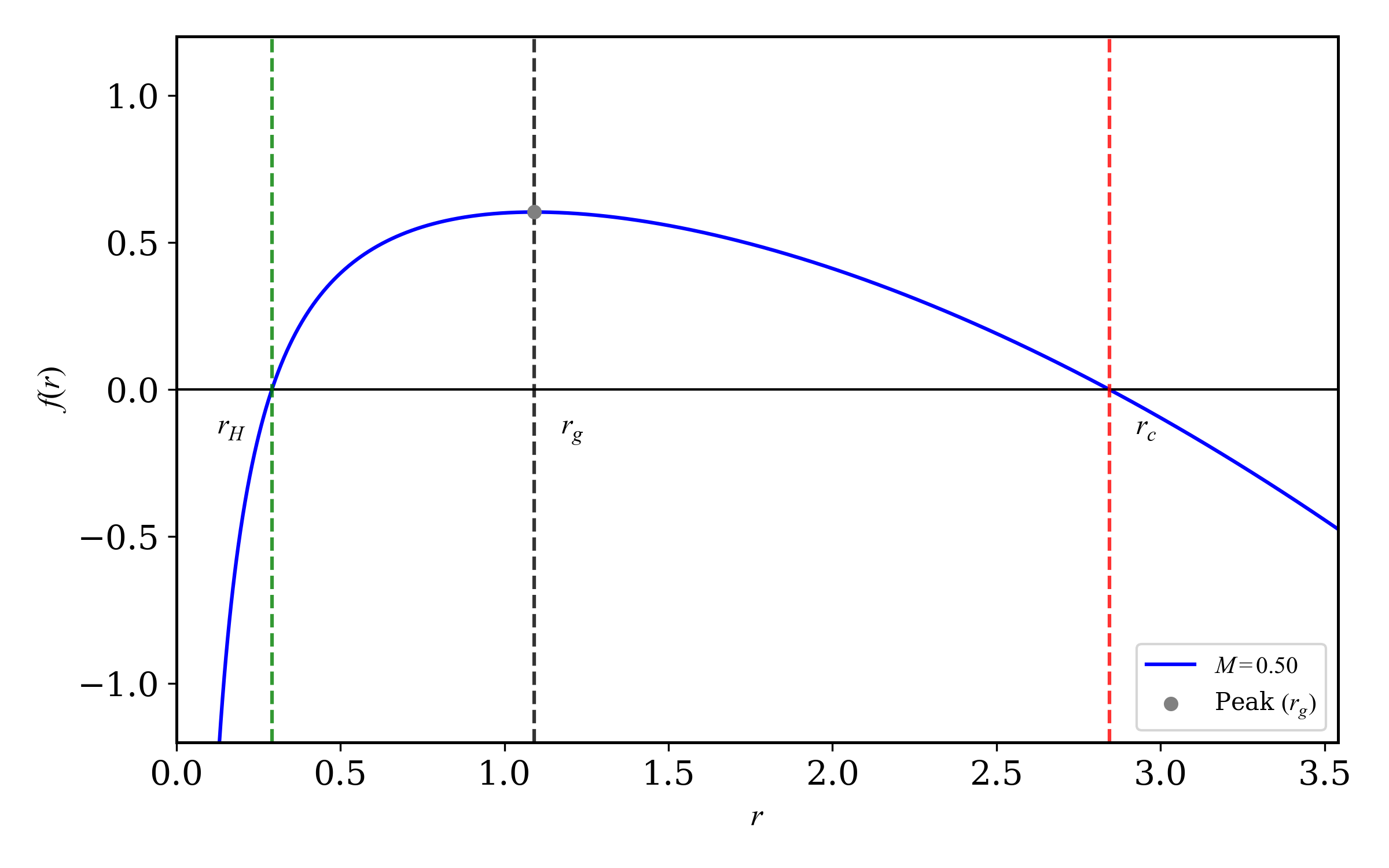}
\caption{{The metric function of the Schwarzschild-de Sitter spacetime. The roots $r=r_H$ and $r=r_c$	
correspond to the black hole event horizon and the cosmological event horizon, respectively. At $r=r_b$, the proper acceleration vanishes, representing the zero-force surface.}\label{Met1}} 
\end{figure}

From \eqref{Ricci}, we can write these equations in the form
\begin{equation}
\label{LHSR}
    \frac{R_{0}^{\ 0}-R_{1}^{\ 1}}{2} - R_{2}^{\ 2} =\frac{\beta'(r) e^{-\beta(r)}}{r} +\frac{1}{r^2}-\frac{e^{-\beta(r)}}{r^2} =\frac{1}{r^2}\left(-\partial_{r}(r e^{-\beta(r)})+1\right).
\end{equation}
Note that, in the case of varnishing of potential, $T_{0}^{\ 0}=T_{1}^{\ 1}$, while $T_{2}^{\ 2}=T_{3}^{\ 3}$ for spherical symmetry, the Einstein field equation gives
\begin{equation}
     \frac{R_{0}^{\ 0}-R_{1}^{\ 1}}{2} - R_{2}^{\ 2} = \kappa^2 T_{0}^{\ 0}.
\end{equation}
By using \eqref{LHSR}, we can write the relation between the metric component and $T_{0}^{\ 0}$ as 
\begin{equation}
    \partial_{r}(r e^{-\beta(r)}) = 1-r^{2}\kappa^{2}T_{0}^{\ 0}.
\end{equation}
Now, we can integrate the above equation to obtain the following relation \cite{LOUSTO1988411}
\begin{equation}
    \label{black react}
    e^{-\beta(r)} = 1-\frac{\kappa^2 M}{r}-\frac{r^2}{l^2}-\frac{\kappa^{2}}{r}\int_{r}^{r_{b}} dr' r'^2 T_{0}^{\ 0}(r'),
\end{equation}
where $r_{b}$ is the upper limit of integration, defined by the condition $f'(r_{b})=0$. This limit possesses physical significance, serving as the critical radius that divides the black hole geometry into two distinct regions \cite{Chang-Young:2010lou}. This behavior is illustrated in Fig.\ref{Met1}, which plots the metric function of the Schwarzschild-de Sitter black hole. The plot clearly demonstrates the separation between the black hole and cosmological horizons. Consequently, $r_b$, given by $r_b^3 = \frac{l^2 \kappa^2 M}{2}$, can be identified as the zero-force surface where the black hole's gravitational attraction is perfectly balanced by cosmic repulsion. Choosing this specific surface as the integration boundary is physically imperative for evaluating the effective field theory corrections. In asymptotically de Sitter spacetimes, there is no spatial infinity to serve as a reference boundary. Integrating the higher-derivative backreaction out to the cosmological horizon $r_c$ would improperly mix the degrees of freedom of the black hole with those of the cosmological background. By terminating the integral at the zero-force surface $r_b$, we effectively enclose the black hole within a quasi-local thermodynamic cavity. This mathematically isolates the black hole's quantum corrections from the background cosmic expansion, ensuring that the perturbative shifts in the metric strictly reflect the intrinsic properties of the localized black hole state.

The De Sitter–Schwarzschild term in the metric arises from the unperturbed action, while the integration terms provide corrections consisting of two contributions. The first contribution arises from the perturbation of the 3-form energy-momentum tensor, while the second originates from higher-order derivatives, which we treat as an effective matter source.

From \eqref{paction}, the equation of motion describing dynamics of 3-form gauge field is given by
\begin{equation}
    \begin{aligned}
       \nabla_{\mu} F^{\mu\nu\sigma\rho} &= {g_{3}}^{2}\Bigg(48 c_4 \nabla_{\mu}(R F^{\mu\nu\sigma\rho}) + 12 c_5 \nabla_{\mu}(R^{\lambda\mu}F_{\lambda}^{\ \nu\sigma\rho}-3 R^{\lambda\rho}F_{\lambda}^{\ \nu\sigma\mu})\\
        &+ 12 c_6 \nabla_{\mu}(R^{\mu\nu\alpha\beta}{F^{\sigma\rho}}_{\alpha\beta}+R^{\mu\rho\alpha\beta}{F^{\nu\sigma}}_{\alpha\beta}+R^{\nu\sigma\alpha\beta}{F^{\mu\rho}}_{\alpha\beta}+R^{\sigma\rho\alpha\beta}{F^{\mu\nu}}_{\alpha\beta})
        \\&+96 c_7 \nabla_{\mu}(F^2 F^{\mu\nu\sigma\rho})+48 c_{8}\nabla_{\mu}(\Box F^{\mu\nu\sigma\rho})\Bigg),
    \end{aligned}
\end{equation}
where $F^2 \equiv F_{\mu\nu\rho\sigma}F^{\mu\nu\rho\sigma}$ and all corrections are treated as a perturbation acting as the source in the 3-form field equation. The perturbed solution of the 3-form field can be calculated using the unperturbed solution from the previous section, and takes the form:
\begin{equation}
    \widetilde{F}^{\mu\nu\sigma\rho} = F^{\mu\nu\sigma\rho}+\Delta F^{\mu\nu\sigma\rho},
\end{equation}
where $\Delta F^{\mu\nu\sigma\rho}$ is the correction. Since the only non-vanishing component is $F^{0123}$, the correction is written as
\begin{equation}
    \label{Fn}
    \begin{aligned}
        \Delta F^{0123} = \frac{24\sqrt{6}{g_{3}}^{3}}{\ell^{3}\kappa^{2}\sqrt{-g}}\bigg(\frac{c_{6}\kappa^{4}\ell^{2}M}{r^{3}}+2\big((12 c_{4}+3 c_{5}+2 c_{6})\kappa^{2}-288 c_{7}{g_{3}}^{2} \big)     \bigg).
    \end{aligned}
\end{equation}
The correction from $c_8$ vanishes due to $\nabla_{\alpha}F^{\mu\nu\sigma\rho}=0$ calculated from the unperturbed metric. Therefore, the perturbed 3-form energy-momentum tensor is in the form
\begin{equation}
    \begin{aligned}
        \Delta \widetilde{T}_{0}^{ \ 0} = \Delta \widetilde{T}_{i}^{ \ i} = -\frac{144 {g_{3}}^{2}}{\ell^{4}\kappa^{4}}\bigg(\frac{c_{6}\kappa^{4}\ell^{2}M}{r^{3}}+2\big((12 c_{4}+3 c_{5}+2 c_{6})\kappa^{2}-288 c_{7}{g_{3}}^{2}   \big)   \bigg).
    \end{aligned}
\end{equation}
This result can be integrated to obtain the first-order contribution to the metric tensor using \eqref{black react}. The metric takes the form:
\begin{equation}
    \label{1stCon}
    \begin{aligned}
        f_{1}(r)=48 {g_{3}}^{2}\Bigg(\frac{3c_{6}\kappa^{2}M}{\ell^{2}r}\ln\left(\frac{r_{b}}{r}\right)+ \frac{2 }{\ell^{4}\kappa^{2}}\bigg(\frac{r_{b}^{3}-r^{3}}{r} \bigg)\bigg(288 c_{7}{g_{3}}^{2} -(12 c_{4}+3 c_{5}+2 c_{6})\kappa^{2}     \bigg).
    \end{aligned}
\end{equation}
The second contribution arises from the perturbed energy-momentum tensor. To find this, we vary the action \eqref{paction} with respect to the metric tensor, keeping terms linear in the constants $c_i$. We define the form of the perturbed energy-momentum tensor as follows:
\begin{equation}
    \Delta T_{\mu}^{\ \nu} = \sum_{i=1}^{8} c_{i}(\Delta T_{i})_{\mu}^{\ \nu},
\end{equation}
where $(\Delta T_{i})_{\mu}^{\ \nu}$ are shown in the appendix \ref{A1}.

By using the unperturbed solution and the relation \eqref{black react}, the second contribution can be found in the form:
\begin{equation}
\begin{aligned}
    f_2(r) = \frac{72 {g_{3}}^{2}}{\ell^{4}\kappa^{2}}\bigg(\frac{\ell^{3}-r^{3}}{r} \bigg)\bigg(288 c_{7}{g_{3}}^{2}-(12 c_{4}+3 c_{5}+2 c_{6})\kappa^{2}      \bigg).
\end{aligned}    
\end{equation}
The full form of the metric with the correction term can be written as
\begin{equation}
\begin{aligned}
\label{metric}
    \widetilde{f}(r) =& f(r)+f_{1}(r)+f_{2}(r) 
    \\=& 1- \frac{\kappa^2 M}{r}-\frac{r^2}{\ell^2}-\frac{24 {g_{3}}^{2}}{\ell^{4}\kappa^{2}}\bigg(\frac{r_{b}^{3}-r^{3}}{r} \bigg)\bigg(288 c_{7}{g_{3}}^{2}  -(12 c_{4}+3 c_{5}+2 c_{6})\kappa^{2}    \bigg)\\
    &+\frac{144c_{6}\kappa^{2}M}{\ell^{2}r}\ln\left(\frac{r_{b}}{r}\right).
\end{aligned}
\end{equation}

\subsection{Black Hole Entropy}
\label{BHE}
Having established the geometrical solution of the black hole supported by a 3-form field, we now turn to the thermodynamic properties of the system. In particular, we analyze the entropy by incorporating higher-order corrections encoded in the coefficients, $c_i$. Our first approach to demonstrating the constraint proceeds by computing the black hole entropy using the Wald formalism \cite{Wald:1993nt,Bodendorfer:2013wga}, which expresses the entropy as a surface integral involving the variation of the Lagrangian with respect to the Riemann tensor. The Wald entropy is given by

\begin{equation}
    \widetilde{S} = - 2\pi \int_{\Sigma} \frac{\delta \widetilde{\mathcal{L}}}{\delta R_{\mu\nu\sigma\rho}}\epsilon_{\mu\nu}\epsilon_{\sigma\rho},
\end{equation}
where $\epsilon_{\mu\nu}$ is binormal to the horizon. In the case of spherical symmetry, the integral becomes 
\begin{equation}
    \widetilde{S} = -2\pi \int_{r=\widetilde{r}_H}r^2 \sin\theta d\theta d\phi \frac{\delta \widetilde{\mathcal{L}}}{\delta R_{\mu\nu\sigma\rho}}\epsilon_{\mu\nu}\epsilon_{\sigma\rho} = -2\pi \widetilde{A}\frac{\delta \widetilde{\mathcal{L}}}{\delta R_{\mu\nu\sigma\rho}}\epsilon_{\mu\nu}\epsilon_{\sigma\rho} \bigg|_{r = \widetilde{r}_H},
\end{equation}
where $\widetilde{r}_H$ is the perturbed event horizon with the surface area $\widetilde{A}$, and all quantities are  evaluated using the perturbed metric. Let the perturbed horizon be expressed in terms of the unperturbed horizon as $\widetilde{r}_{H} = r_{H} + \Delta r_{H}$ and the surface area $\widetilde{A} = A+\Delta A$. Under the expansion of the Lagrangian $\widetilde{\mathcal{L}} =\mathcal{L}+\Delta \mathcal{L} $, the perturbed entropy becomes
\begin{equation}
    \widetilde{S} = -2\pi \left(A\frac{\delta\mathcal{L}}{\delta R_{\mu\nu\sigma\rho}}+A \frac{\Delta\mathcal{L}}{\delta R_{\mu\nu\sigma\rho}}+\Delta A\frac{\delta\widetilde{\mathcal{L}}}{\delta R_{\mu\nu\sigma\rho}}+...     \right)\epsilon_{\mu\nu}\epsilon_{\sigma\rho}\bigg|_{r = \widetilde{r}_H}.
\end{equation}
The first term corresponds to the unperturbed entropy, since the variation gives: 
\begin{equation}
   S= -2\pi A\frac{\delta\mathcal{L}}{\delta R_{\mu\nu\sigma\rho}}\epsilon_{\mu\nu}\epsilon_{\sigma\rho} = -\frac{\pi A}{\kappa^2}\widetilde{g}^{\mu\rho}\widetilde{g}^{\nu\sigma}\epsilon_{\mu\nu}\epsilon_{\sigma\rho} = \frac{2\pi A}{\kappa^2}.
\end{equation}
In the last step, we used the identity for the binormal tensor, $\epsilon_{\mu\nu}\epsilon^{\mu\nu} = -2$.
By keeping only the first-order terms in the perturbation, the entropy can be rewritten as
\begin{equation}
\label{Ds}
    \Delta S = \widetilde{S}-S = \Delta S_{1}+\Delta S_{2},
\end{equation}
where 
\begin{equation}
\label{s1}
    \Delta S_{1} = -2\pi A \frac{\delta \Delta\mathcal{L}}{\delta R_{\mu\nu\sigma\rho}}\epsilon_{\mu\nu}\epsilon_{\sigma\rho}\bigg|_{g_{\mu\nu},r_{H}}
\end{equation}
and
\begin{equation}
\label{s2}
    \Delta S_{2} = -2\pi\Delta A\frac{\delta\mathcal{L}}{\delta R_{\mu\nu\sigma\rho}} = \frac{2 \pi}{\kappa^2}\Delta A.
\end{equation}

To evaluate the entropy in \eqref{Ds}, we start with equation \eqref{s1}, where the variation is given by:
\begin{equation}
\label{dls1}
\begin{aligned}
     \frac{\delta \Delta \mathcal{L}}{\delta R_{\mu\nu\rho\sigma}} = &
2 c_1 R g^{\mu\rho} g^{\nu\sigma} 
+ 2 c_2 R^{\mu\rho} g^{\nu\sigma} 
+ 2 c_3 R^{\mu\nu\rho\sigma}+\frac{c_4}{2}(F^{2}g^{\mu\sigma}g^{\nu\rho}-F^{2}g^{\mu\rho}g^{\nu\sigma})\\
&+c_5 F^{\alpha\beta\gamma[\mu}F_{\alpha\beta\gamma}^{\ \ \ \ \nu}g^{\sigma\rho]}+c_6(F^{\alpha\beta\mu[\sigma|}F_{\alpha\beta}^{\ \ \nu|\rho]}+F^{\mu\nu\alpha\beta}F^{\sigma\rho}_{\ \ \ \alpha\beta}).
\end{aligned}
\end{equation}
By using the unperturbed metric to evaluate \eqref{dls1} and substituting it into \eqref{s1}, the result is
\begin{equation}
    \Delta S_{1} = \frac{2\pi A}{\kappa^{2}}\Bigg[\frac{8\kappa^{2}c_{3}(3-\eta)}{\ell^{2}(1-\eta)}+ 12 \frac{\kappa^{2}}{\ell^{2}}(4 c_{1}+c_{2}) - \frac{24}{\ell^{2}}(12 c_{4}+3c_{5}+2c_{6}){g_{3}}^{2}    \Bigg],
\end{equation}
where we define new parameter, $\eta$, expressed as
\begin{equation}
    \eta = 1 -\frac{2 r_H^3}{l^2 \kappa^2 M}.
\end{equation}
Note that the parameter $\eta$ vanishes at the extremal limit.
For \eqref{s2}, we can find $\Delta A$ by expanding the perturbed metric to the first order, as follows  
\begin{equation}
    \begin{aligned}
      0 =  \widetilde{f}(r_H+\Delta r_H) =& \widetilde{f}(r_H)+ \partial_{r}\widetilde{f}(r_{H})\Delta r_H\\
      =& f(r_{H})+\Delta f(r_H) + \partial_{r}\widetilde{f}(r_{H})\Delta r_H.
    \end{aligned}
\end{equation}
Considering only the first-order perturbation, we find 
\begin{equation}
    \label{drH}
    \Delta r_H = -\frac{\Delta f(r_H)}{\partial_{r}f(r_H)}.
\end{equation}
This leads to the perturbed horizon area, given by
\begin{equation}
    \label{dA}
    \Delta A = \widetilde{A}-A = 8\pi r_H \Delta r_H = -\frac{8\pi r_H \Delta f(r_H)}{\partial_r f(r_H)}.
\end{equation}
Substituting this into \eqref{s2}, the result becomes
\begin{equation}
\begin{aligned}
     \Delta S_{2} =& \frac{96 {g_{3}}^{2} \pi A}{\kappa^{6}\ell^{4}M\eta}\Bigg[\frac{1}{2} \big(288 c_{7}{g_{3}}^{2}-(12 c_{4}+ 3 c_{5}+2 c_{6})\kappa^{2}\big)\big(2 r_{b}^{3} -  \ell^{2}\kappa^{2}M  (1-\eta)\big)\\
     &- 2 c_{6}\kappa^{4}\ln{\left(\frac{2 r_{b}^{3}}{\ell^{2}\kappa^{2}M(1-\eta)}    \right)}  \Bigg]
\end{aligned}
\end{equation}
Therefore, the entropy in \eqref{Ds} becomes
\begin{equation}
    \label{Dsf}
    \begin{aligned}
       \Delta S =& \frac{S}{\eta}\Bigg[\frac{8\kappa^{2}c_{3}\eta(3-\eta)}{\ell^{2}(1-\eta)}+ 12 \kappa^{2}\eta(4 c_{1}+c_{2}) - \frac{24\eta}{\ell^{2}}(12 c_{4}+3c_{5}+2c_{6}){g_{3}}^{2} \\
       &\quad\quad\quad+\left(\frac{24{g_{3}}^{2}}{\kappa^{4}M\ell^{4}}\right)\bigg( \big(288 c_{7}{g_{3}}^{2}-(12 c_{4}+ 3 c_{5}+2 c_{6})\kappa^{2}\big)\big(2 r_{b}^{3}-
        \ell^{2}\kappa^{2} M (1-\eta)\big)\\
     &\quad\quad\quad- \frac{4 c_{6}\kappa^{4}}{\ell^{2}}\ln{\left(\frac{2 r_{b}^{3}}{\ell^{2}\kappa^{2}M(1-\eta)}    \right)} \bigg) \Bigg].
    \end{aligned}
\end{equation}
The entropy is required to be invariant under field redefinitions, as demonstrated in section \ref{sec:redef}. To satisfy this condition, the integration limit $r_{b}$ is chosen to be $r_{b}^{3} = \frac{\ell^{2}\kappa^{2}M}{2}$. The invariant entropy is then given by
\begin{equation}
\label{DS}
\begin{aligned}
      \Delta S = &S\Bigg[\frac{8 c_{3} \kappa^{2}}{\ell^{2}}\left(\frac{3-\eta}{1-\eta}\right) + \frac{12\kappa^{2}}{\ell^{2}}\big(4 c_{1}+c_{2} \big) +\frac{48{g_{3}^{2}}}{\kappa^{2}\ell^{2}} \big(144 c_{7}{g_{3}}^{2} - (12 c_{4}+3 c_{5}+ 2 c_{6})\kappa^{2}\big) \\
      &+ 96 c_{6} {g_{3}}^{2} \frac{\ln{(1-\eta})}{\ell^{2}\eta}  \Bigg].
\end{aligned}
\end{equation}
According to the thermodynamic analyses presented in \cite{Chang-Young:2010lou}, the black hole geometry is divided into two independent systems, partitioned by a zero-temperature wall at $r=r_b$. It is required that, for a fixed mass and cosmological constant, the corrected entropy continues to satisfy standard thermodynamic principles at the event horizon:
\begin{equation}
    \Delta S > 0,
\end{equation}
This condition leads to the inequality
\begin{equation}
    \label{newBound}
    \begin{aligned}
         &2 c_{3} \kappa^{2}\left(\frac{3-\eta}{1-\eta}\right) + 3\kappa^{2}\big(4 c_{1}+c_{2} \big) +\frac{12{g_{3}^{2}}}{\kappa^{2}} \big(144 c_{7}{g_{3}}^{2} - (12 c_{4}+3 c_{5}+ 2 c_{6})\kappa^{2}\big)\\
         &+24 c_{6} {g_{3}}^{2} \frac{\ln{(1-\eta})}{\eta} > 0 .
    \end{aligned}
\end{equation}
The above bound represents consistency conditions of the action with higher derivatives and therefore remains valid irrespective of the background.

The inequality \eqref{newBound} in the limit $\eta = 0$ reduces to
\begin{equation}
\label{sbound0}
    2 c_{3} \kappa^{2} + \kappa^{2}\big(4 c_{1}+c_{2} \big) +\frac{4{g_{3}^{2}}}{\kappa^{2}} \big(144 c_{7}{g_{3}}^{2} - (12 c_{4}+3 c_{5}+ 4 c_{6})\kappa^{2}\big) > 0 .
\end{equation}
This bound remains invariant under the field redefinitions outlined in section \ref{sec:redef}. Given the restriction to four-dimensional spacetime in this analysis, the coefficients $c_{1}, c_{2}$ and $c_{3}$ correspond to topological invariants and may be disregarded. As a result, the thermodynamic positivity bound is reduced to
\begin{equation}
    144 c_{7}{g_{3}}^{2} - (12 c_{4}+3 c_{5}+ 4 c_{6})\kappa^{2} > 0. \label{eq:thermo-positivity-bound}
\end{equation}
This bound imposes background-independent constraints on the coupling constants in the effective action \eqref{paction}.

In the case of an extremal black hole, the correction to the condition $z=\frac{3\sqrt{3}\kappa^{2}M}{2\ell} = 1$ can be obtained by introducing a shifted ratio $\tilde{z} = z+\Delta z$. The horizon condition for the corrected metric component is
\begin{equation}
    0 = \tilde{f}(\tilde{r}_{H},z) = f(r_{H},z)+\Delta f(r_{H},z)+\Delta r_{H}\partial_{r_{H}}f(r_{H},z)+\Delta z \partial_{z}f(r_{H},z),
\end{equation}
where the first and third terms vanish due to extremality. The resulting expression for the extremal shift is
\begin{equation}
    \Delta z = - \frac{\Delta f(r_{H},z)}{\partial_{z}f(r_{H},z)}.
\end{equation}
A direct evaluation gives
\begin{equation}
    \Delta z =0.
\end{equation}

This result implies that there is no shift in the extremal mass parameter, leading to $\Delta M = 0$. This striking feature arises directly from the physics of the quasi-local thermodynamic boundary, $r_b$. As the black hole approaches the extremal limit, the event horizon $r_H$, the cosmological horizon $r_c$, and the zero-force surface $r_b$ all converge to a single degenerate radius, characterizing the well-known Nariai limit.

Consequently, the spatial volume of the thermodynamic cavity—over which the higher-derivative backreaction is integrated—shrinks to exactly zero. Without an extended spatial domain for the effective energy density of the quantum corrections to accumulate, their net backreaction on the metric identically vanishes at this limit.

It is worth contrasting this with the asymptotically flat case, where higher-derivative terms typically induce a downward shift of the extremal mass and so permit the kinematic decay demanded by the Weak Gravity Conjecture. The comparison should not be pushed too far here. The 3-form flux is not an independent charge: by Eq. (\ref{eq:lambda}) it fixes the cosmological constant itself, so there is no charge-to-mass ratio in this theory for the corrections to shift, and the extremality condition (\ref{b1}) is the Nariai bound rather than a mass-versus-charge statement. What we find is simply that the correction to the extremal mass vanishes. Because the quasi-local cavity volume goes to zero in the degenerate-horizon limit, the exact Nariai state saturates the classical bound instead of being pushed below it. This is a property of the degenerate-horizon limit of the cavity, not a topological one.

To observe a dynamic mass shift driven by thermodynamic positivity, one must consider the near-extremal regime where the cavity volume remains finite ($\eta > 0$). This rigidity underscores the subtlety of formulating Swampland criteria in spacetimes with multiple competing horizons, confirming that the zero-force surface $r_b$ is not merely a mathematical regulator, but a necessary physical boundary that correctly captures the phase transition of the Nariai geometry.

\subsection{Field-Redefinition Invariance of the Bound}
\label{sec:redef}
 
Next, we would like to address a question that
determines whether the bound~\eqref{sbound0} constrains the theory at all.
In a higher-derivative effective field theory the individual couplings
$c_i$ are not physical observables: a redefinition of the metric
reshuffles them, trading one operator for another without altering any
on-shell quantity. A constraint written in terms of specific $c_i$ could
therefore be an artifact of the chosen field frame rather than a genuine
statement about the theory. This is precisely the concern that a bound
derived from black-hole data might merely re-express the extremality
condition on the mass. To settle it, we now show that the entropy
correction, and hence the bound, can be written entirely in terms of
field-redefinition-invariant combinations of the couplings.
 
Under a field redefinition, the metric transforms as
\begin{equation}
g_{\mu\nu} \rightarrow g_{\mu\nu} + \delta g_{\mu\nu} + \mathcal{O}(\delta g^2),
\end{equation}
from which the resulting shift in the action can be derived as follows:
\begin{equation}
\delta S = \int d^{4}x \sqrt{-g}\Bigg[\frac{1}{2\kappa^{2}}\delta g^{\mu\nu}\bigg(R_{\mu\nu} - \frac{1}{2}g_{\mu\nu}R - \kappa^{2} T_{\mu\nu} \bigg)\Bigg],
\end{equation}
or, in terms of the Lagrangian density,
\begin{equation}
\delta\mathcal{L} = \frac{1}{2\kappa^{2}}\delta g^{\mu\nu}\bigg(R_{\mu\nu} - \frac{1}{2}g_{\mu\nu}R - \kappa^{2} T_{\mu\nu} \bigg).
\end{equation}
Let $\delta g_{\mu\nu}$ be a general perturbation up to second order in derivatives:
\begin{equation}
\delta g_{\mu\nu} = r_{1}R_{\mu\nu} + r_{2} g_{\mu\nu} R + r_{3} \kappa^{2} g_{\mu\nu} F^{2},
\end{equation}
where the coefficients $r_{i}$ are arbitrary constants. The
energy-momentum tensor of the unperturbed 3-form gauge field simplifies to
\begin{equation}
T_{\mu\nu} = \frac{g_{\mu\nu}}{48 {g_{3}}^{2}}F^{2}.
\end{equation}
The shift in the Lagrangian then becomes
\begin{equation}
\delta\mathcal{L} = \frac{1}{2\kappa^{2}}\bigg( r_{1}R_{\mu\nu} + r_{2} g_{\mu\nu} R + r_{3} \kappa^{2} g_{\mu\nu} F^{2}\bigg)\bigg( R^{\mu\nu} - \frac{1}{2}g^{\mu\nu}R - \frac{\kappa^{2} g^{\mu\nu}}{48 {g_{3}}^{2}}F^{2}\bigg),
\end{equation}
which yields
\begin{equation}
\begin{aligned}
\delta\mathcal{L} = &\frac{1}{2\kappa^{2}}\bigg(\frac{r_{1}}{2}+r_{2} \bigg)R^{2} - \frac{r_{1}}{2\kappa^{2}}R_{\mu\nu}R^{\mu\nu}+\bigg(-\frac{r_{1}}{96 {g_{3}}^{2}}+ \frac{r_{2}}{12 {g_{3}}^{2}}+\frac{r_{3}}{2 \kappa^{2}}\bigg)R F^{2} \\
&+\frac{r_{1}}{12 {g_{3}}^{2}}R^{\mu\nu}{F_{\mu}}^{\sigma\rho\delta}{F_{\nu\sigma\rho\delta}}+\frac{\kappa^{2}r_{3}}{24 {g_{3}}^{2}}F^{4}.
\end{aligned}
\end{equation}
Comparing this with the effective action in Eq.~\eqref{paction}, given by
\begin{equation}
S = \int d^{4}x\sqrt{-g}\Bigg(\frac{R}{2 \kappa^{2}} - \frac{1}{48 {g_{3}}^{2}}F^{2} + \delta\mathcal{L} \Bigg),
\end{equation}
we deduce the following set of transformations for the coefficients:
\begin{equation}
\begin{aligned}
\Delta c_{1} &= \frac{1}{2\kappa^{2}}\bigg(\frac{r_{1}}{2}+r_{2} \bigg) &
\Delta c_{2} &= - \frac{r_{1}}{2\kappa^{2}} &
\Delta c_{3} &= 0 &
\Delta c_{4} &= -\frac{r_{1}}{96 {g_{3}}^{2}}+ \frac{r_{2}}{12 {g_{3}}^{2}}+\frac{r_{3}}{2 \kappa^{2}}\\
\Delta c_{5} &= \frac{r_{1}}{12 {g_{3}}^{2}} &
\Delta c_{6} &= 0 &
\Delta c_{7} &= \frac{r_{3}}{24 {g_{3}^{2}}} &
\Delta c_{8} &= 0.
\end{aligned}
\end{equation}
It is evident that $c_3, c_6$ and $c_8$ are explicitly invariant under
these field redefinitions. We can construct another invariant combination
by requiring
\begin{equation}
144\Delta c_{7} {g_{3}}^{4} - (12 \Delta c_{4} + 3 \Delta c_{5}){g_{3}}^{2}\kappa^{2} + \frac{1}{4}(4 \Delta c_{1}+\Delta c_{2})\kappa^{4} = 0,
\end{equation}
which provides the invariant quantity:
\begin{equation}
c_{\text{inv}} = 144 c_{7} {g_{3}}^{4} - (12 c_{4} + 3  c_{5}){g_{3}}^{2}\kappa^{2} + \frac{1}{4}(4  c_{1}+ c_{2})\kappa^{4}.
\end{equation}
Consequently, the entropy correction in Eq.~\eqref{DS} can be rewritten in
terms of the invariant quantities $c_3$, $c_6$ and $c_{\text{inv}}$ as
\begin{equation}
\label{eq:DS-invariant}
\Delta S = S\left[\frac{8c_3\kappa^{2}}{\ell^{2}}\left(\frac{3-\eta}{1-\eta}\right)
+\frac{48\,c_{\text{inv}}}{\kappa^{2}\ell^{2}}
+\frac{96\,c_6 {g_3}^{2}}{\ell^{2}}\left(\frac{\ln(1-\eta)}{\eta}-1\right)\right].
\end{equation}
Since $c_3$, $c_6$ and $c_{\text{inv}}$ are each invariant under the
redefinitions above, the entropy correction is manifestly independent of
the field frame, and so is the positivity bound that follows from it. In
four dimensions $c_1$, $c_2$ and $c_3$ multiply topological invariants and
drop out, so the physical content of the bound resides entirely in
$c_{\text{inv}}$ and $c_6$. This invariance is what we regard as the
decisive evidence that the constraint is a genuine statement about the
effective theory rather than a re-expression of the extremality condition
on the mass in a particular field frame.

\subsection{Energy condition}

\begin{figure}[htbp]
\centering
\includegraphics[scale=0.8]{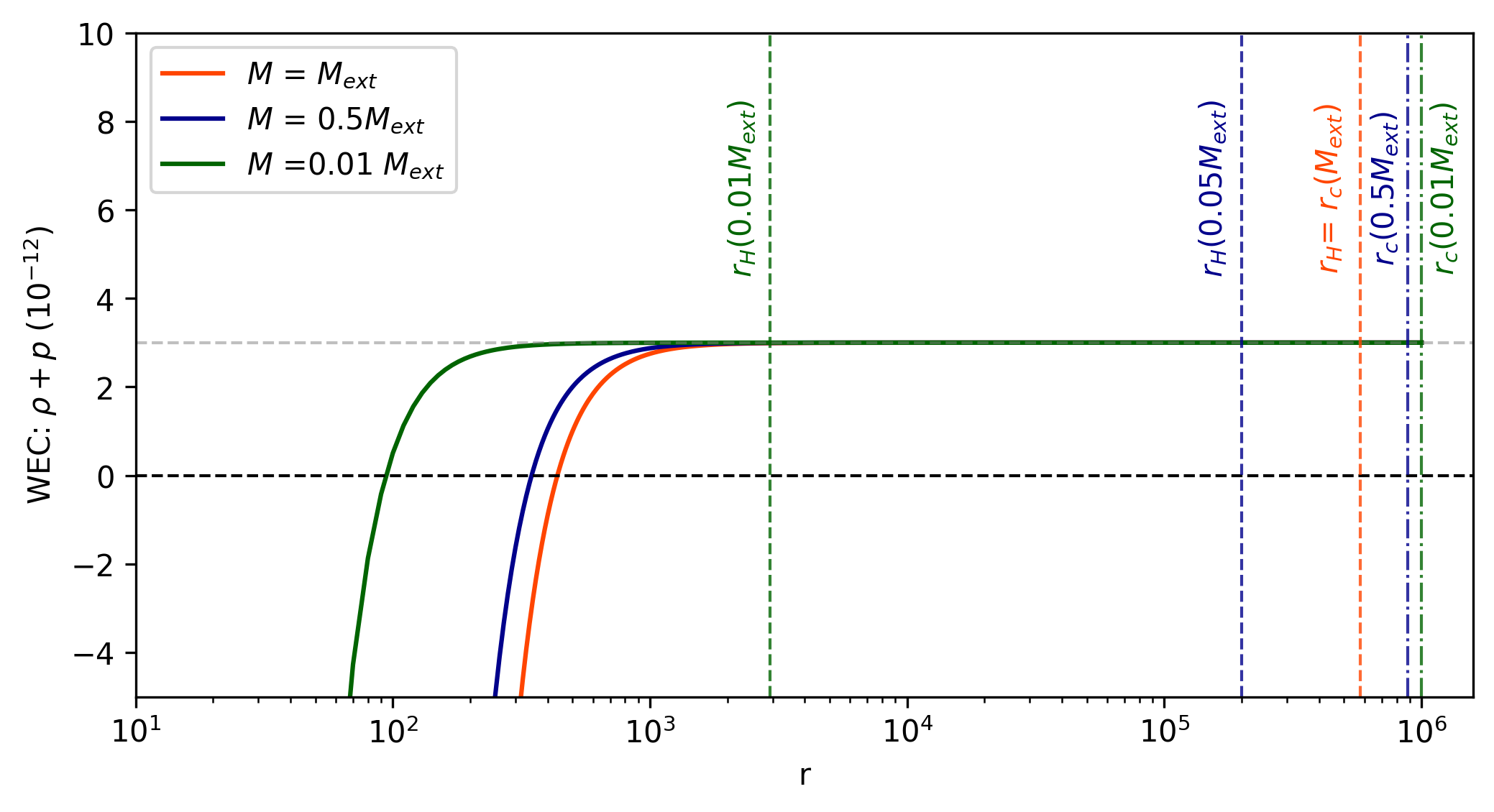}
\caption{Radial profile of the weak energy condition (WEC), expressed as \(\rho + p\), for different mass values: \(M = M_{\text{ext}}\) (orange), \(0.5 M_{\text{ext}}\) (blue), and \(0.01 M_{\text{ext}}\) (green). The solid lines show that the WEC is satisfied (\(\rho + p \geq 0\)) in the regions of interest. Vertical dashed lines indicate the event horizons \(r_H\), while dash-dotted lines represent the corresponding cosmological horizons \(r_c\). The WEC holds between the horizons for all considered masses.\label{WEC}} 
\end{figure}

\begin{figure}[htbp]
\centering
\includegraphics[scale=0.8]{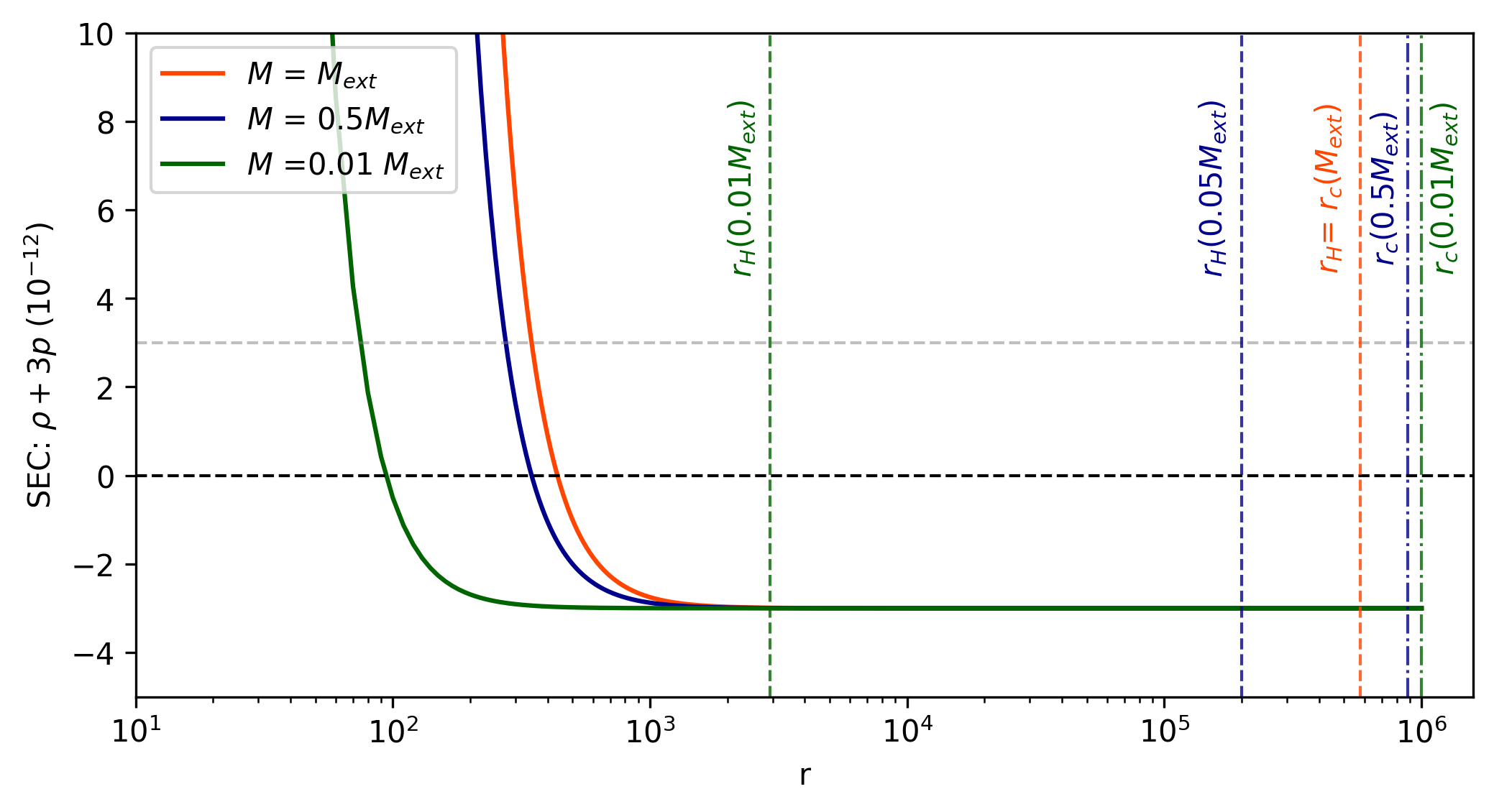}
\caption{Radial profile of the strong energy condition (SEC), expressed as \(\rho + 3p\), for different mass values: \(M = M_{\text{ext}}\) (orange), \(0.5 M_{\text{ext}}\) (blue), and \(0.01 M_{\text{ext}}\) (green). The curves indicate that the SEC is violated (\(\rho + 3p < 0\)) in the region near the black hole for all mass configurations. Vertical dashed lines denote the event horizons \(r_H\), while dash-dotted lines indicate the corresponding cosmological horizons \(r_c\).\label{SEC}} 
\end{figure}

In this subsection, we evaluate the energy conditions of the model. The energy density and pressure are written as
\begin{equation}
\begin{aligned}
     T_{0}^{\ 0} &= - \rho =- \left(\frac{3}{\kappa^{2}l^{2}}+\frac{3{g_{3}}^{2}}{\kappa^{4}l^{4}}\bigg((144 c_4 + 36 c_5 +48 c_{6})\kappa^{2} - 6912 {g_{3}}^{2}c_{7}   \bigg) + \frac{144 c_{6}M}{l^{2}r^{3}}\right)\\
      T_{i}^{\ j} &=  p\delta_{i}^{\ j}=- \left(\frac{3}{\kappa^{2}l^{2}}+\frac{3{g_{3}}^{2}}{\kappa^{4}l^{4}}\bigg((144 c_4 + 36 c_5 +48 c_{6})\kappa^{2} - 6912 {g_{3}}^{2} c_{7}   \bigg) + \frac{144 c_{6}M}{l^{2}r^{3}}\right)\delta_{i}^{\ j} \label{eq:EnergyConditions}
\end{aligned}
\end{equation}
The standard energy conditions in general relativity include:
\begin{itemize}
    \item \textbf{Null Energy Condition (NEC)}: \(\rho+p \geq 0\)
    \item \textbf{Weak Energy Condition (WEC)}: \(\rho \geq 0\) and \(\rho + p \geq 0\)
    \item \textbf{Strong Energy Condition (SEC)}: \(\rho + p \geq 0\) and \(\rho + 3p \geq 0\)
\end{itemize}
For a large dS radius, $l \sim m_p/\sqrt{\Lambda} \sim \left(10^6 \;{\rm eV}^{-1}\right) m_p $, the second term in the energy density and pressure is strongly suppressed. As a result, energy conditions are independent from the higher-derivative corrections. From the above condition, it is clear that the NEC is always valid, while the WEC imposes a constraint on the energy density ($\rho \geq 0$). The WEC is violated near $r\rightarrow 0$  and leads to the cosmological constant $\frac{3}{\kappa^{2}l^{2}}$, which corresponds to a large dS radius. The behavior of WEC is shown in Figure \ref{WEC}. In the case of SEC, the second condition reduces to $-2\rho \geq 0$, which implies $\rho \leq 0$. This violates the WEC shown in Figure \ref{SEC}, which requires the energy density to be non-negative.

While the Weak Energy Condition is satisfied in the exterior region between the event and cosmological horizons, it is violated in the limit $r \rightarrow 0$. However, this breakdown region lies entirely inside the event horizon ($r < r_H$) and is strictly causally disconnected from the exterior spacetime. Therefore, this central WEC violation does not compromise the thermodynamic positivity bounds---which are explicitly evaluated at the event horizon---nor does it affect the validity of the energy conditions in the physically accessible exterior domain, where the effective field theory remains robust and under perturbative control.

\section{Cosmological Inflation}
\label{Cosmosec}

In this section, we study cosmological inflation using the action \eqref{paction} to explore how the bound \eqref{eq:thermo-positivity-bound}, originally derived from black hole thermodynamics, constrains an expanding universe. Specifically, we analyze the dynamics of inflationary solutions driven by a 3-form gauge field and examine whether the positivity condition for entropy corrections remains valid in this cosmological setting.

Before proceeding with the inflationary dynamics, it is crucial to contextualize this approach within the broader Swampland program. The original de Sitter Swampland Conjecture strongly disfavors stable de Sitter vacua by bounding the slope of the scalar potential, demanding $m_P \frac{|\nabla V|}{V} \ge c \sim \mathcal{O}(1)$ \cite{Obied:2018sgi}. This condition is famously in tension with the extremely flat potentials required for standard slow-roll inflation \cite{Agrawal:2018own}. However, the Refined de Sitter Conjecture introduces a crucial caveat, permitting transient de Sitter-like phases provided the potential is sufficiently tachyonic, satisfying $m_P^2 \frac{\min(\nabla_i \nabla_j V)}{V} \le -c' \sim -\mathcal{O}(1)$ \cite{Ooguri:2018wrx}. This refinement provides a broad theoretical pathway for certain inflationary scenarios---particularly those exploring plateau-like or concave-down regimes where the second slow-roll parameter is inherently negative ($\eta_\chi < 0$)---to alleviate the strict Swampland tension. While achieving a phenomenologically viable number of e-folds often still necessitates $c' \ll 1$ in such frameworks, our primary objective is not to fully resolve this $\mathcal{O}(1)$ debate. Rather, we treat the thermodynamic positivity bound derived in Section 3---which originates from the requirement of a well-defined UV completion for black hole entropy---as a separate, complementary constraint. By mapping our 3-form action to an effective scalar potential, we aim to determine how this independent, UV-derived thermodynamic bound restricts the available parameter space for slow-roll inflation in the IR.

The simplest approach to modeling inflation with a 3-form gauge field is to consider its 4-form field strength tensor as being Hodge dual to a scalar field. This duality allows us to express the 4-form as
\begin{equation}
\label{Fdual}
F_{\mu\nu\sigma\rho} = \lambda \epsilon_{\mu\nu\sigma\rho} \Phi,
\end{equation}
where $\lambda$ is a constant and $\Phi$ plays the role of an effective inflaton field. By substituting Eq. (\ref{Fdual}) into the higher-derivative action, this duality acts as the exact physical bridge, translating the 3-form couplings directly into the non-minimal conformal factor and the effective scalar potential.

It is important to note, however, that the assumptions imposed on the 3-form field at cosmological and black hole scales differ significantly due to the symmetries of each background. In the cosmological case, we assume a homogeneous and time-dependent field configuration compatible with the symmetries of a Friedmann–Lemaître–Robertson–Walker (FLRW) universe. In contrast, the black hole analysis relies on a static, spherically symmetric ansatz. While these configurations might appear contradictory, they are valid within their respective regimes of applicability and are not expected to interfere with each other, as they describe physics at vastly different scales. Nonetheless, for consistency in our effective field theory, we assume that the higher-order coefficients $c_i$ appearing in the action are universal and apply across both setups, allowing the thermodynamic positivity bounds to directly constrain the inflationary parameter space.

In the context of cosmological inflation, the metric takes the form  
\begin{equation}
    \label{cosmetric}
    ds^2 = -dt^2 + a(t)^2\left( dx^2 + dy^2 + dz^2 \right).
\end{equation}
Substituting \eqref{Fdual} and \eqref{cosmetric} into the generalized action \eqref{paction}, we obtain the simplified form  
\begin{equation}
\begin{aligned}
    S = \int d^4x \sqrt{-g} \bigg[ &\left(1 - \frac{4(12 c_4 + 3c_5 + 2c_6)}{m_P^2} \lambda^2 \Phi^2 \right)\left( \frac{\dot{a}^2}{a^2} + \frac{\ddot{a}}{a} \right) \frac{6m_P^2}{2} \\
    & - 24 \lambda^2 c_8\, \partial^\mu \Phi\, \partial_\mu \Phi 
    + \frac{1}{2} \frac{\lambda^2}{{g_{3}}^{2}} \Phi^2 
    + 576 c_7 \lambda^4 \Phi^4 \bigg],
\end{aligned}
\end{equation}
where \( m_P^2 = \frac{1}{\kappa^2} \) is the reduced Planck mass squared. The first term corresponds to the Ricci scalar in the FLRW background,  
\begin{equation}
    R = 6\left( \frac{\dot{a}^2}{a^2} + \frac{\ddot{a}}{a} \right),
\end{equation}
non-minimally coupled to the scalar field \(\Phi\). The second term gives the kinetic energy, while the remaining terms represent the potential, including contributions from both the original 3-form field (\(F^2\)) and higher-order corrections.

To move to the physical (Einstein) frame, we rescale the metric by a conformal factor \(\Omega(\Phi)\):  
\begin{equation}
    d\hat{s}^2 = \Omega^2(\Phi) ds^2 = - \Omega^2(\Phi) dt^2 + \Omega^2(\Phi) a(t)^2 (dx^2 + dy^2 + dz^2).
\end{equation}
In this frame, the metric becomes
\begin{equation}
    d\hat{s}^2 = -d\hat{t}^2 + \hat{a}(\hat{t})^2 \delta_{ij} dx^i dx^j,
\end{equation}
where the new time and scale factor are defined via
\begin{equation}
    d\hat{t} = \Omega(\Phi)\, dt, \quad \hat{a}(\hat{t}) = \Omega(\Phi)\, a(t).
\end{equation}

The Ricci scalar transforms under the conformal transformation as
\begin{equation}
    \label{conformalR}
    R = \Omega^2 \left[ \hat{R} + 6 \hat{\Box} \ln \Omega - 6 \hat{g}^{\mu\nu} (\partial_\mu \ln \Omega)(\partial_\nu \ln \Omega) \right],
\end{equation}
where the second term can be eliminated by integration by parts. To simplify the coupling between the scalar field and gravity, the conformal factor is typically chosen such that
\begin{equation}
\label{Omega}
    \Omega^2(\Phi) = 1 - \mu\frac{\lambda^2\Phi^2}{m_{P}^2},
\end{equation}
where $\mu = 4(12 c_4 + 3c_5+2c_6)$. After the conformal transformation, the action in the Einstein frame becomes
\begin{equation}
\begin{aligned}
    \hat{S} = \int d^4 x \sqrt{-\hat{g}}\bigg[&\frac{m_{P}}{2}\hat{R}-\frac{\hat{g}^{\mu\nu}}{2}\left(\frac{48\lambda^2 c_8}{\Omega^2}+\frac{6\mu^2\lambda^4 \Phi^2}{\Omega^4 m_{P}^2}\right)\partial_{\mu}\Phi\partial_{\nu}\Phi 
    \\&+\frac{1}{\Omega^4}\left( \frac{1}{2}\frac{\lambda^2}{{g_{3}}^{2}}\Phi^2+ 576 c_7 \lambda^4 \Phi^4\right)\bigg].
\end{aligned}
\end{equation}
Moreover, we redefine the scalar field $\Phi$ as $\chi$ using
\begin{equation}
\label{chi}
   \frac{ d\chi}{d\Phi} = \pm \sqrt{\frac{48\lambda^2 c_8}{\Omega^2}+ \frac{6\mu^2\lambda^4 \Phi^2}{\Omega^4 m_{P}^2}},
\end{equation}
to transform the action into the canonical form
\begin{equation}
\label{actionnew}
      \hat{S} = \int d^4 x \sqrt{-\hat{g}}\bigg[\frac{\hat{R}}{2 m_{P}^2}-\frac{\hat{g}^{\mu\nu}}{2}\partial_{\mu}\chi\partial_{\nu}\chi-V(\chi)      \bigg],
\end{equation}
where we define the potential as 
\begin{equation}
\label{Vchi}
V(\chi) = -\frac{1}{\Omega^4}\left( \frac{1}{2}\frac{\lambda^2}{{g_{3}}^{2}}\Phi^2 + 576 c_7 \lambda^4 \Phi^4\right).
\end{equation}

In the small-field limit, we assume \( \frac{6\mu^2\lambda^2\Phi^2}{m_{P}^2} \ll 48 c_8 \), allowing us to simplify equation \eqref{chi} to
\begin{equation}
\label{chismall}
    \chi = \pm \lambda \sqrt{48 c_{8}} \, \Phi.
\end{equation}
Substituting into the potential \eqref{Vchi}, we obtain
\begin{equation}
    V(\chi) = -\frac{\chi^{2}}{96 {g_{3}}^{2}c_{8}} - \frac{c_{7} \chi^{4}}{4 c_{8}^{2}},
\end{equation}
where the conformal factor simplifies to \( \Omega^{2} = 1 \). The potential has minima at
\[
\chi = \pm \frac{1}{4\sqrt{3}{g_{3}}} \sqrt{\frac{c_{8}}{|c_{7}|}},
\]
when $c_7 < 0$. The corresponding vacuum energy at the minimum is negative:
\[
V\left(\pm \frac{1}{4\sqrt{3}{g_{3}}} \sqrt{\frac{c_{8}}{|c_{7}|}}\right) = -\frac{1}{9216 |c_{7}| {g_{3}}^{4}}.
\]
Since this vacuum corresponds to an anti-de Sitter space, we will not explore the small-field regime further in this work.

In the large field limit, where \( \frac{6\mu^2\lambda^2\Phi^2}{m_{P}^2} \gg 48 c_8 \), the equation \eqref{chi} simplifies to
\begin{equation}
    \frac{d\chi}{d\Phi} = \pm \frac{\sqrt{6}\mu\lambda^{2}\Phi}{\Omega^{2} m_{P}} \sqrt{1 - \frac{48 c_{8}}{6\mu}}.
\end{equation}
To avoid an imaginary conformal factor, we assume \( \mu < 0 \). Under this condition, the solution becomes
\begin{equation}
\label{chilarge}
    \pm\frac{\chi}{m_{P}} = \sqrt{ \frac{3}{2} \left( 1 + \frac{48 c_{8}}{|\mu|} \right) } \ln \left( 1 + |\mu| \lambda^2 \frac{\Phi^2}{m_{P}^2} \right).
\end{equation}
In the regime of large field inflation, where \( \left| \frac{ \sqrt{|\mu|} \lambda \Phi }{ m_{P} } \right| \gg 1 \), we can invert Eq.~\eqref{chilarge} to express \(\Phi\) in terms of \(\chi\):
\begin{equation}
\label{phi2}
    \Phi^2 = \frac{m_{P}^2}{|\mu|\lambda^2} \exp \left( \sqrt{ \frac{2}{3\gamma} } \frac{|\chi|}{m_{P}} \right),
\end{equation}
where \( \gamma \equiv 1 + \frac{8 c_{8}}{|\mu|} \). The absolute value of \(\chi\) is used to ensure consistency with the large field assumption. Using Eqs.~\eqref{Omega} and \eqref{phi2}, the potential can be written as
\begin{equation}
    V(\chi) = -m_{P}^4 \left( \frac{1}{2|\mu| {g_{3}}^{2}m_{P}^2} + \frac{576 c_7}{|\mu|^2} e^{\sqrt{ \frac{2}{3\gamma} } \frac{|\chi|}{m_{P}}} \right)
    \left( e^{- \sqrt{ \frac{1}{6\gamma} } \frac{|\chi|}{m_{P}} } + e^{ \sqrt{ \frac{1}{6\gamma} } \frac{|\chi|}{m_{P}} } \right)^{-2}. \label{eq:potential}
\end{equation}
The equation of motion for \(\chi(\hat{t})\), derived from the action \eqref{actionnew}, is
\begin{equation}
\label{eqX}
    \ddot{\chi} + 3 H \dot{\chi} + \frac{dV}{d\chi} = 0.
\end{equation}
Under the slow-roll approximation, the standard slow-roll parameters are defined as
\begin{align}
    \epsilon_{\chi} &= \frac{m_{P}^2}{2} \left( \frac{V'(\chi)}{V(\chi)} \right)^2\label{epsix}, \\
    \eta_{\chi} &= m_{P}^2 \left( \frac{V''(\chi)}{V(\chi)} \right)\label{etax}.
\end{align}
The number of e-folds is then given by
\begin{equation}
\label{efold}
    N_{\chi} = \frac{1}{m_{P}^2} \int_{\chi_e}^{\chi_i} d\chi \, \frac{V(\chi)}{V'(\chi)},
\end{equation}
where \(\chi_i\) and \(\chi_e\) denote the values of the scalar field at the beginning and end of inflation, respectively, with the end point defined by the condition \( \epsilon_{\chi} = 1 \).

From the slow-roll parameters, one can derive the spectral index and tensor-to-scalar ratio:
\begin{align}
     n_s &= 1 - 6\epsilon_{\chi} + 2\eta_{\chi}, \\
     r &= 16 \epsilon_{\chi}.
\end{align}
These predictions can be evaluated at the horizon exit (typically at \( N_{\chi} \sim 40 - 60 \)) to compare with current observational constraints from cosmic microwave background (CMB) data, such as those provided by Planck.

\subsection{Results}
To analyze the parameter space suitable for slow-roll inflation, we substitute the explicit functional form of the potential \eqref{eq:potential} into the slow-roll definitions. This yields exact analytical expressions for the slow-roll parameters, $\epsilon_{\chi}$ and $\eta_{\chi}$, along with the scalar power spectrum, $P_s$ (The detailed algebraic derivations are provided in Appendix \ref{app:tensor_scalar_ratio}). To evaluate the inflationary dynamics, the final field value, $\chi_e$, is determined by the condition $\epsilon_{\chi}(\chi_e) = 1$, giving
\begin{equation}
\epsilon_{\chi} = \frac{\left((2304 c_{7}{g_{3}}^{2}m_{P}^2 - \mu)\exp\left[\sqrt{\frac{2}{3\gamma}} \frac{|\chi|}{m_{P}}\right] + \mu \right)^2}{3\gamma \left(1 + \exp\left[\sqrt{\frac{2}{3\gamma}} \frac{|\chi|}{m_{P}}\right]\right)^2 \left(1152 c_{7}{g_{3}}^{2}m_{P}^2 \exp\left[\sqrt{\frac{2}{3\gamma}} \frac{|\chi|}{m_{P}}\right] + \mu \right)^2},
\end{equation}
\begin{equation}
    \eta_\chi = \frac{2 \left( - (2304c_7 g_3^2 m_P^2 - \mu) \exp\left[2\sqrt{\frac{2}{3\gamma}}\frac{|\chi|}{m_P}\right] + 4(1152c_7 g_3^2 m_P^2 - \mu) \exp\left[\sqrt{\frac{2}{3\gamma}}\frac{|\chi|}{m_P}\right] + \mu \right)}{3\gamma \left(1 + \exp\left[\sqrt{\frac{2}{3\gamma}}\frac{|\chi|}{m_P}\right]\right)^2 \left(1152c_7 g_3^2 m_P^2 \exp\left[\sqrt{\frac{2}{3\gamma}}\frac{|\chi|}{m_P}\right] + \mu\right)},
\end{equation}
and
\begin{equation}
    P_s = -\frac{\gamma}{16\pi^2 \mu^2 g_3^2 m_P^2} \frac{\exp\left[\sqrt{\frac{2}{3\gamma}}\frac{|\chi|}{m_P}\right] \left( 1152c_7 g_3^2 m_P^2 \exp\left[\sqrt{\frac{2}{3\gamma}}\frac{|\chi|}{m_P}\right] + \mu \right)^3}{\left( (2304c_7 g_3^2 m_P^2 - \mu)\exp\left[\sqrt{\frac{2}{3\gamma}}\frac{|\chi|}{m_P}\right] + \mu \right)^2}.\label{eq:powerspectrum}
\end{equation}

The e-folding number is given by
\begin{equation}
\begin{aligned}
N_{\chi}(\chi) =& \sqrt{\frac{3\gamma}{2}} \frac{\chi}{m_{P}} + \frac{1728 \gamma c_{7} {g_{3}}^{2}m_{P}^{2} e^{\sqrt{\frac{2}{3\gamma}} \frac{\chi}{m_{P}}}}{2304 c_{7} {g_{3}}^{2} m_{P}^{2} - \mu}\\
&- \frac{3\gamma(1152 c_{7}{g_{3}}^{2}m_{P}^{2} - \mu)^{2} \ln{\left[e^{\sqrt{\frac{2}{3\gamma}} \frac{\chi}{m_{P}}}(2304 m_{P}^{2} - \mu) + \mu\right]}}{(2304 c_{7} {g_{3}}^{2} m_{P}^{2} - \mu)^{2}}.
\end{aligned}
\end{equation}

To facilitate the analysis, we consider the case where \(\mu < 0\) and fix \(c_8 = \frac{1}{48}\) to simplify the large-field limit. We also define a new parameter that combines the couplings \(c_4\) and \(c_5\) to align with the black hole bound:
\begin{equation}
\bar{c} = 12 c_4 + 3 c_5,
\end{equation}
so that \(\mu\) becomes
\begin{equation}
|\mu| = |4 \bar{c} + 8 c_6|.
\end{equation}

The vacuum energy is evaluated at \(\chi = 0\), yielding
\begin{equation}
V_0 = V(0) = -\frac{m_{P}^4}{4} \left(\frac{576 c_7}{|\mu|^2} + \frac{1}{2|\mu| {g_{3}}^{2}m_{P}^2}\right).
\end{equation}
For a de Sitter vacuum (\(V_0 > 0\)), the parameter \(c_7\) must be negative. This condition leads to the inequality
\begin{equation}
\label{dSB}
288 |c_7|{g_{3}}^{2} - |\bar{c} + 2 c_6| > 0. 
\end{equation}

The overall amplitude of the potential is not fixed by the constraints derived above. Both
terms in the vacuum energy (\ref{eq:potential}), the quartic contribution $576c_7/|\mu|^2$ and the
kinetic contribution $1/(2|\mu|g_3^2 m_P^2)$, rescale as $V \to s\,V$ under
\begin{equation}
c_7 \;\to\; s\,c_7 , \qquad g_3 \;\to\; g_3/\sqrt{s} , \label{eq:scaling}
\end{equation}
with all other couplings held fixed. This rescaling leaves the invariant combination
$c_7 g_3^2$ and the parameter $\mu$ unchanged, and therefore preserves the thermodynamic
positivity bound (\ref{eq:thermo-positivity-bound}), the de Sitter condition (\ref{dSB}), and the shape observables $n_s$,
$r$ and $N_\chi$: it acts purely on the height of the inflationary plateau. The scale $s$ is
then fixed a posteriori by normalizing the scalar power spectrum (\ref{eq:powerspectrum}) to the observed
value $P_s \simeq 2.1\times10^{-9}$ at horizon exit, which for our benchmark couplings
requires $s \simeq 10^{-8}$. This places the plateau at $V \sim 10^{-8} m_P^4$ and the
inflationary Hubble scale at $H \sim 10^{-4} m_P$, consistent with the estimate used in
Eq.~(\ref{eq:scaling}).



\begin{figure}[htbp]
    \centering
    \begin{subfigure}[b]{0.48\textwidth}
        \centering
        \includegraphics[width=\textwidth]{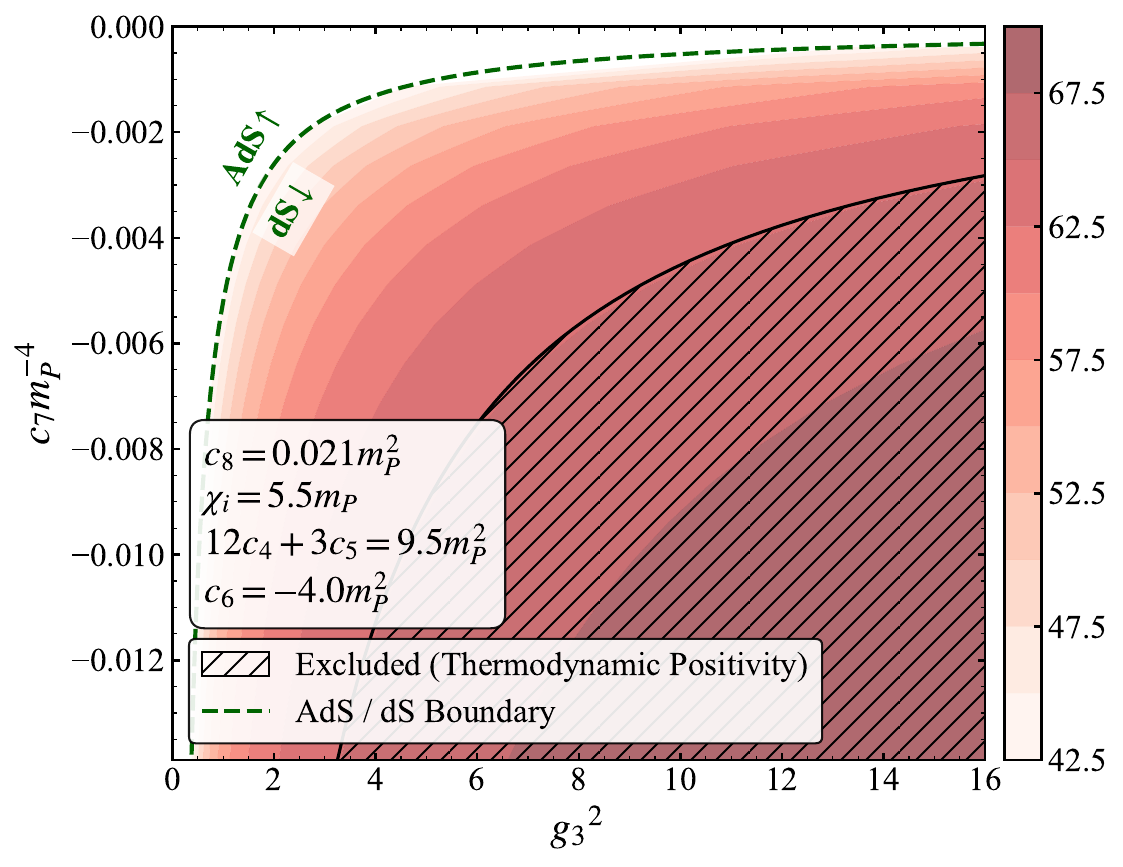}
        \caption{}
        \label{fig:c7vsg3}
    \end{subfigure}
    \hfill
    \begin{subfigure}[b]{0.48\textwidth}
        \centering
        \includegraphics[width=\textwidth]{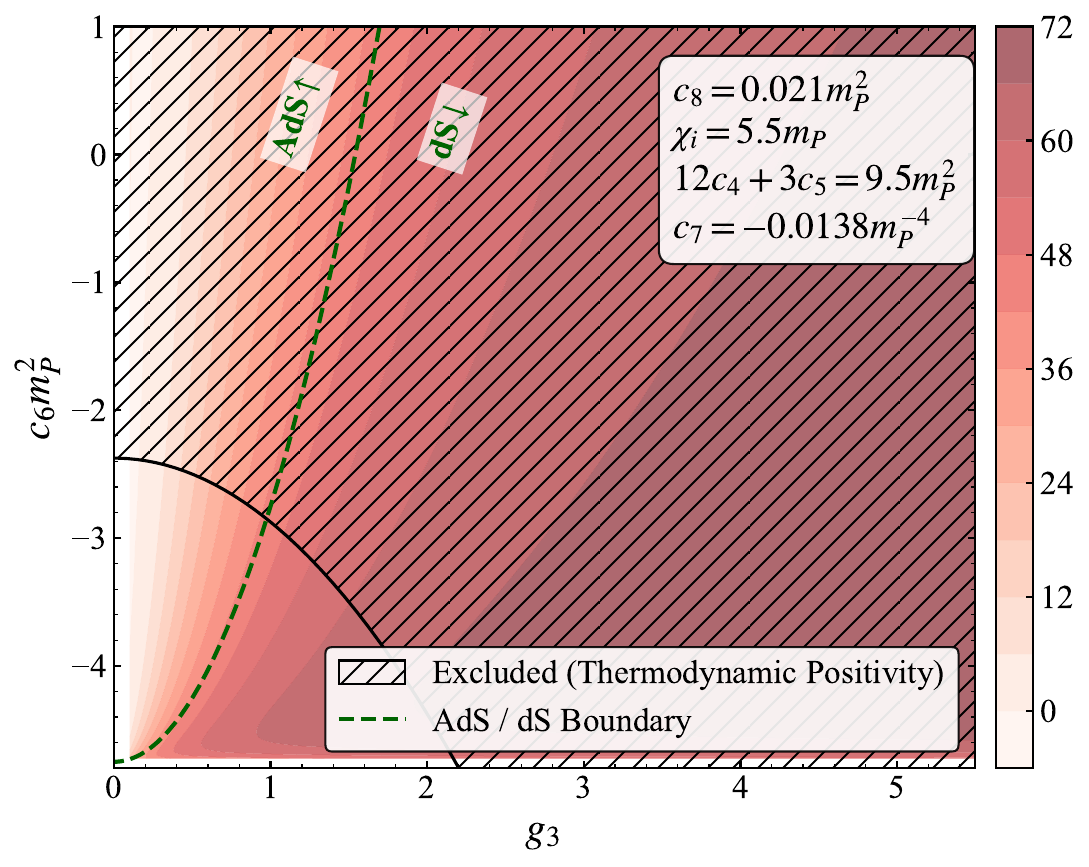}
        \caption{}
        \label{fig:c6vsg3}
    \end{subfigure}
    \hfill
    \begin{subfigure}[b]{0.48\textwidth}
        \centering
         \includegraphics[width=\textwidth]{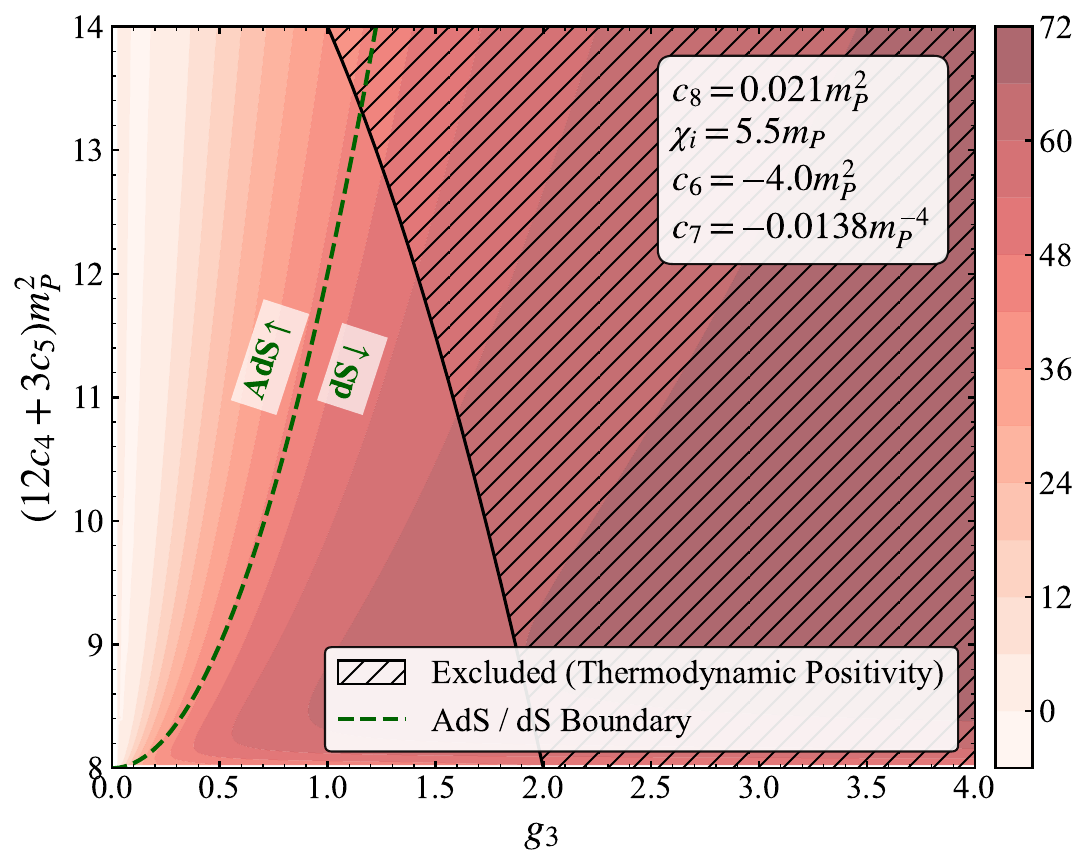}
        \caption{}
        \label{fig:cvsg3}
    \end{subfigure}
      \hfill
    \begin{subfigure}[b]{0.48\textwidth}
        \centering
        \includegraphics[width=\textwidth]{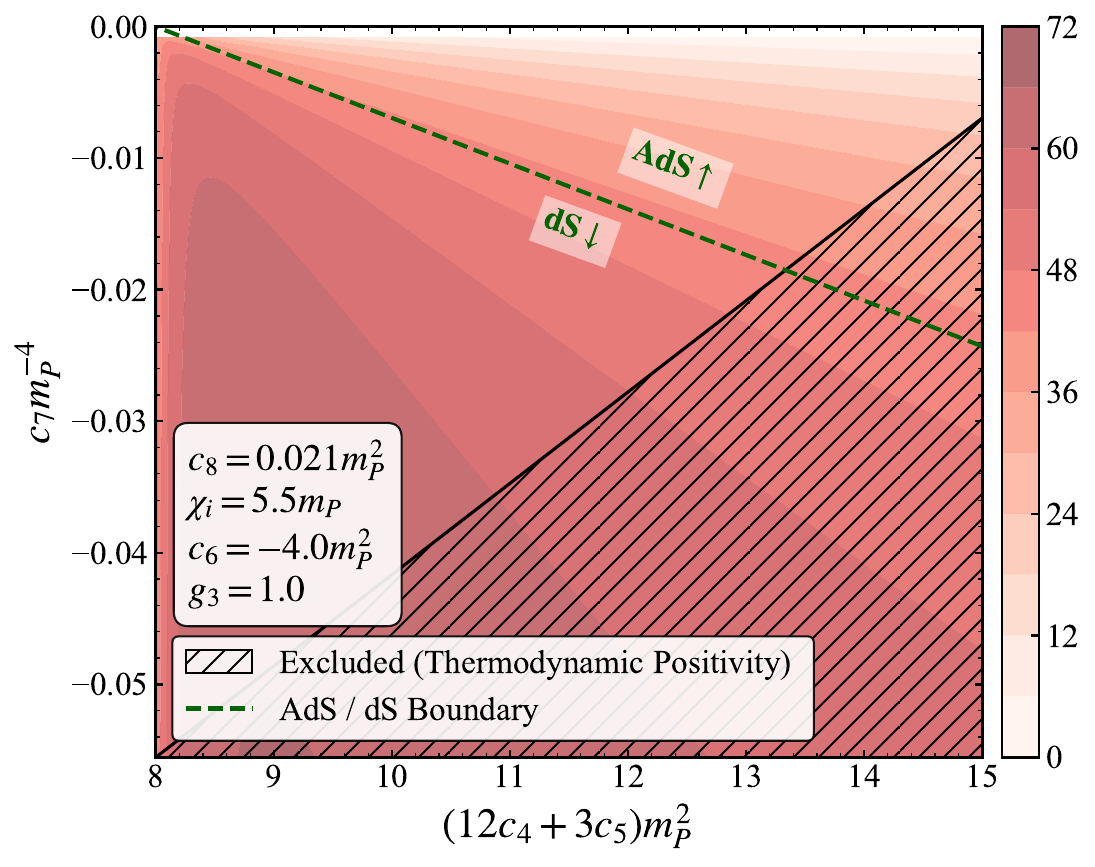}
        \caption{}
        \label{fig:c7vsc}
    \end{subfigure}
    \hfill
    \begin{subfigure}[b]{0.48\textwidth}
        \centering
        \includegraphics[width=\textwidth]{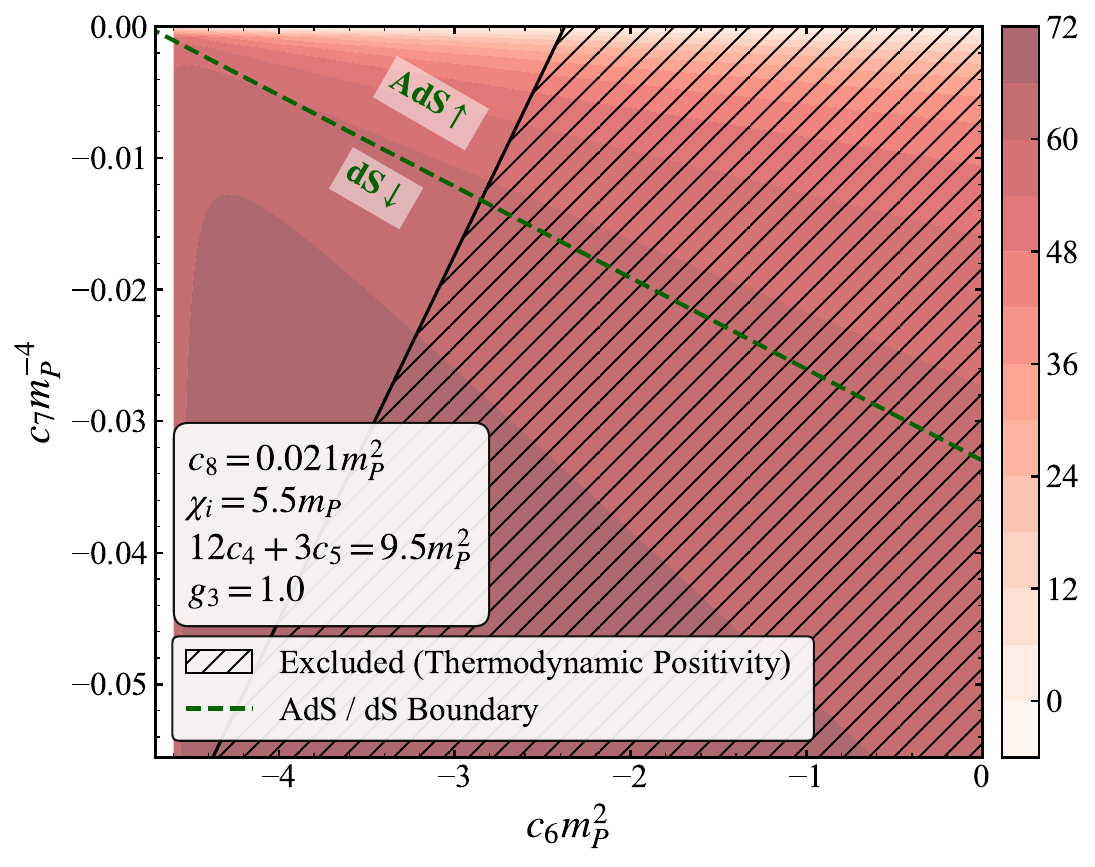}
        \caption{}
        \label{fig:c6vsg3}
    \end{subfigure}
      \hfill
    \begin{subfigure}[b]{0.48\textwidth}
        \centering
        \includegraphics[width=\textwidth]{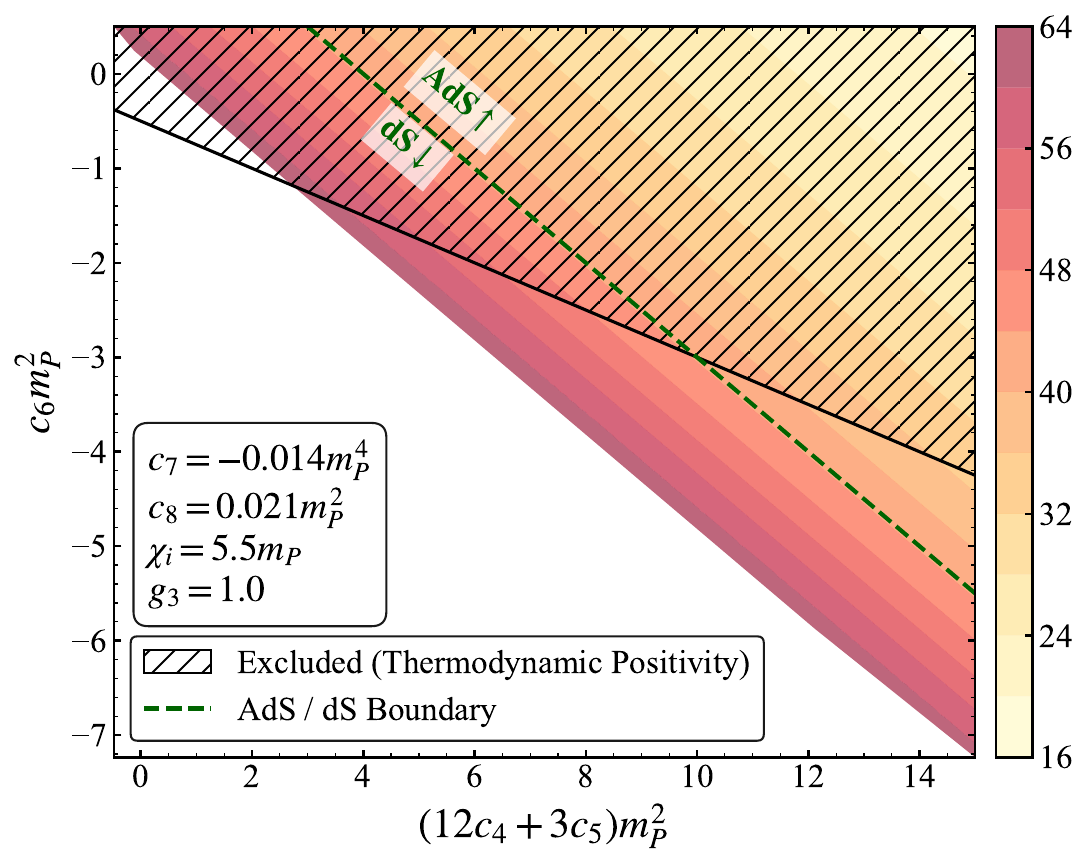}
        \caption{}
        \label{fig:c6vsc}
    \end{subfigure}
    \caption{Contour plots of the e-folding number $N_{\chi}$ in the parameter space of the 3-form coupling $g_{3}$ and higher-derivative couplings. Panel (a) shows the plot of $c_{7}$ versus $g_{3}^{2}$. Panels (b)--(f) illustrate the dependence of $N_{\chi}$ across different combinations of $c_{7}$, $c_{6}$, and $(12c_{4} + 3c_{5})$, with the remaining parameters and the initial field value $\chi_{i}$ fixed as indicated in each inset. The hatched regions denote the parameter space excluded by the thermodynamic positivity bound, while the green dashed lines mark the boundary separating the de Sitter (dS) and anti-de Sitter (AdS) vacua.\label{contourP}}
\end{figure}

\begin{figure}[hbt!]
\centering
\includegraphics[scale=0.58]{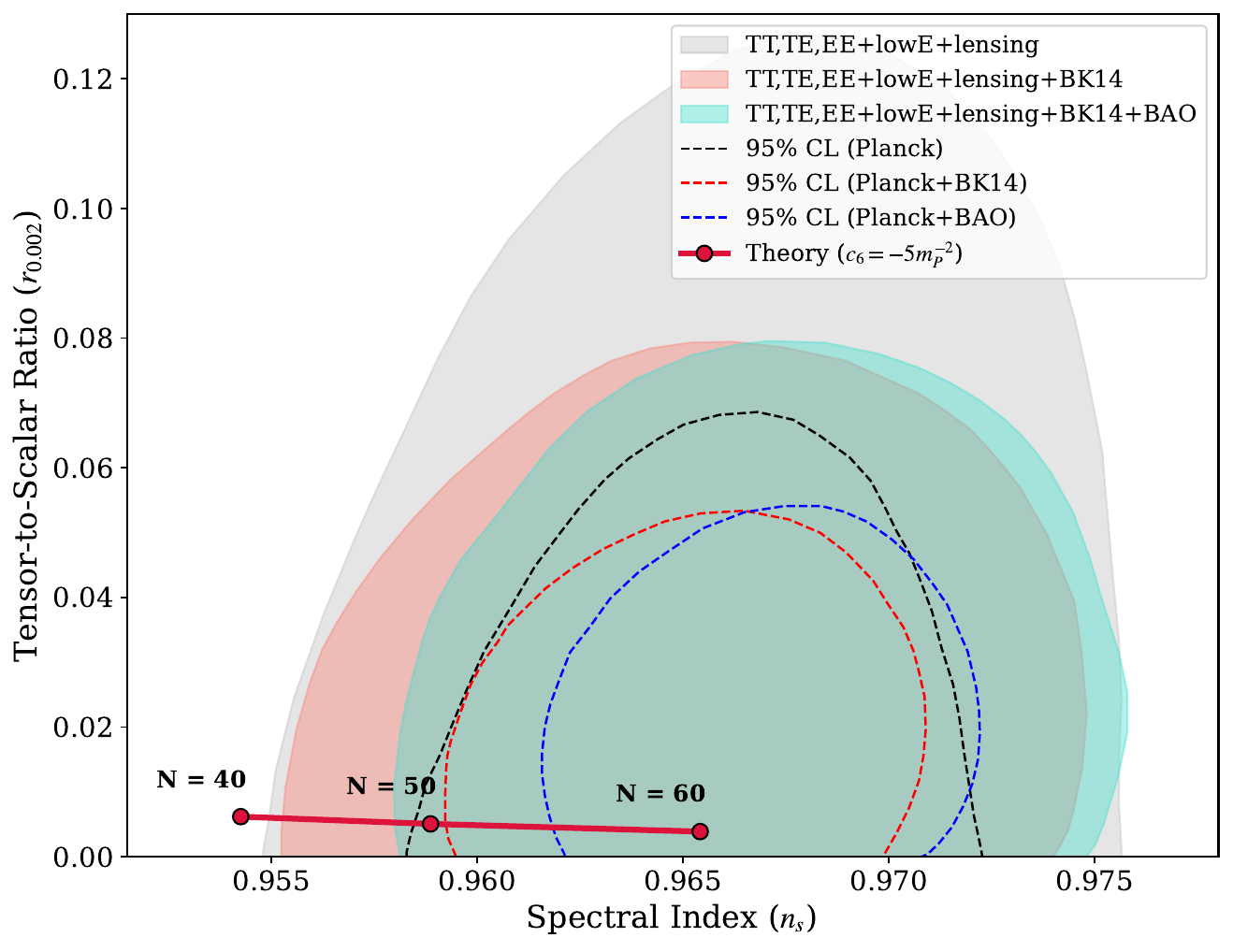}
\caption{Tensor-to-scalar ratio \(r_{0.002}\) and spectral index \(n_s\) for the model with fixed \(c_6 = -5 m_{P}^{-2}\), $c_{7} = -0.014 m_{P}^{4}$, and $g_{3} = 1.0$. The solid green line represents the theoretical predictions for different e-folding numbers $40$, $50$, and $60$. The model predictions fall well within the observationally allowed region, consistent with Planck constraints. \label{ns}}
\end{figure}

\begin{figure}[hbt!]
\centering
\includegraphics[scale=0.7]{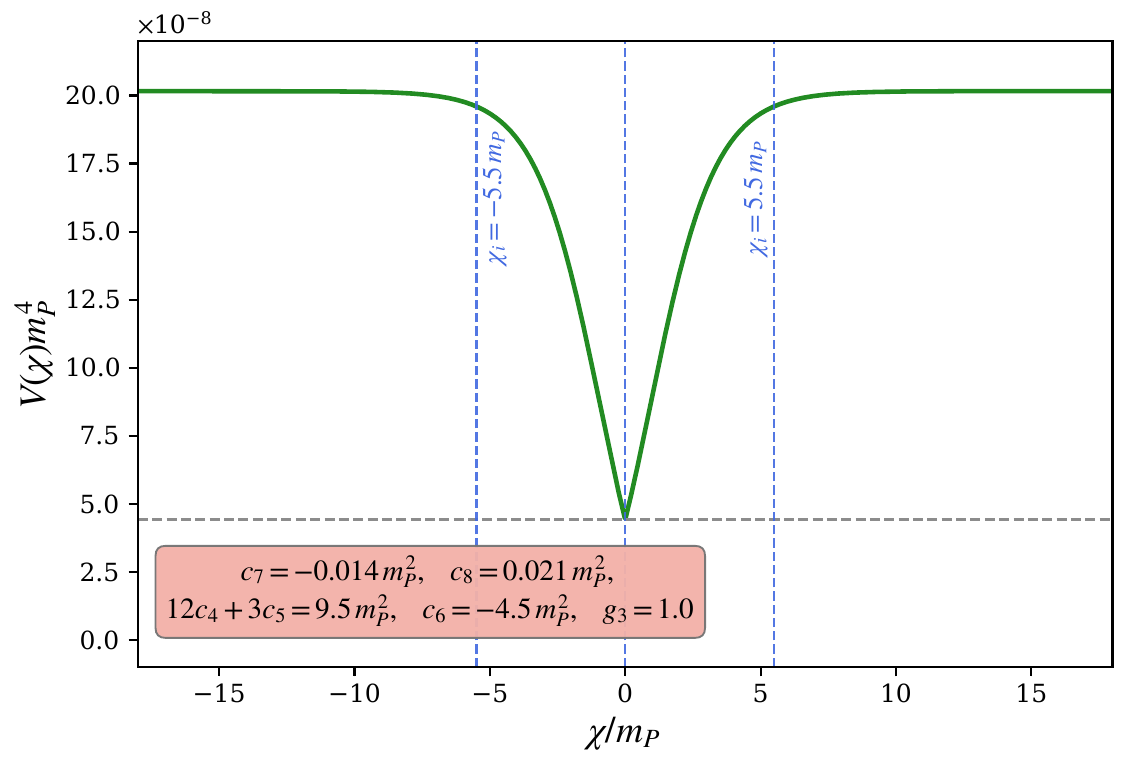}
\caption{Scalar potential $V(\chi)$ as a function of the canonical field $\chi$ plotted for the chosen parameters $c_{7}$, $c_{8}$, $c_{6}$ and combination $12 c_4 + 3 c_5$. The solid green line represents the potential profile, exhibiting a Higgs-like shape characteristic of large-field inflation. The vertical dashed blue lines mark the initial conditions $\chi_{i}\sim \pm 5.5 m_{P}$, while the minimum near $\chi_{e}\sim 0$ corresponds to the end of inflation, defined by $\epsilon_{\chi} = 1$
.\label{potential2}}
\end{figure}

The initial field value $\chi_i$ significantly affects the calculated e-folding number and the viable parameter space. To satisfy the slow-roll condition $\epsilon_{\chi} \ll 1$ while maintaining a consistent dS vacuum, we choose the representative value $\chi_i \approx 5.5 m_{P}$ for our primary analysis. The corresponding contour plots of the e-folding number under this initial condition, evaluated alongside the thermodynamic positivity bound, are comprehensively presented in Figure \ref{contourP}. Panels (a)–(c) illustrate the dependence of the e-folding number on the 3-form coupling $g_3$ while panels (d)–(f) show its variation with respect to the internal combinations of the higher-derivative couplings $c_{7}$, $c_{6}$, and $12 c_4 + 3 c_5$.

From Figure \ref{contourP} (a)--(c), we find that the 3-form coupling $g_{3}$ plays a crucial role in determining the inflationary dynamics when mapped against the higher-derivative parameters. Their variation strongly influences the e-folding number $N_{\chi}$. The allowed range consistent with the theoretical bounds forms a narrow band; for instance, as observed in panel (a), increasing $g_{3}$ requires a correspondingly more negative value of $c_{7}$ to maintain viable slow-roll conditions. The rescaling introduced above corresponds to motion along the curves $c_7g_3^2 = \text{const}$ in the $(c_7, g_3^2)$ plane of panel (a), along which every bound and every contour shown is constant; the structure displayed in Figure 5 is therefore independent of the normalization.

By fixing $g_{3}$ alongside the initial field value $\chi_{i}$, we further examine the direct interplay among the higher-derivative couplings in Figure \ref{contourP} (d)--(f). The contour plots demonstrate that the thermodynamic positivity bound confines inflation to a specific region of the higher-derivative parameter space. Within this specific intersection, the model naturally yields an e-folding number $N_{\chi} \in [44, 64]$, ensuring compatibility with observational requirements. Selecting a representative point from Figure~\ref{fig:c6vsc} and plotting the corresponding potential yields Figure~\ref{potential2}, where the potential exhibits a Higgs-like form and the inflationary trajectory follows a pattern similar to that discussed in~\cite{Mishra:2018dtg}. Because our large-field limit yields a concave-down, Higgs-like potential ($\eta_\chi < 0$), the inflationary dynamics naturally align with the tachyonic condition of the Refined de Sitter Conjecture discussed above.

Using the parameter choices from Figure \ref{fig:c6vsc}, we compute the spectral index \(n_s\) and tensor-to-scalar ratio \(r\), and compare them with Planck data \cite{Planck:2018jri}. As illustrated in Figure \ref{ns}, the theoretical predictions for our model align perfectly with the Planck 2018 constraints. For a benchmark value of $N_{\chi} = 60$, the model explicitly yields $n_{s}\approx 0.966$ and $r_{0.002}\approx 0.0032$, which lies comfortably within the $95\%$ CL region. We find that \((n_s, r)\) in our model is predominantly sensitive to the value of the e-folding number. As shown in Figure \ref{ns}, fixing \(c_6\) and varying combination $12 c_4 + 3 c_5$ demonstrates that the model with \(N_{\chi} \sim 50\!-\!60\) falls entirely within the Planck constraint region, provided the positivity bound is satisfied.

Due to the behavior of the energy conditions at small $r$, the validity of omitting the local $r^{-3}$ correction in Eq.~(\ref{eq:EnergyConditions}) during inflation warrants brief justification. Within the global inflationary regime, the relevant physical scale is the Hubble radius, $r \sim l = H^{-1}$. Evaluating the ratio of the geometric correction to the background energy density at this scale gives $\rho_{\text{corr}}/\rho_{\text{bg}} \approx 48 c_6 \kappa^2 M / l^3$. Applying the extremal mass bound $\kappa^2 M \leq 2l / (3\sqrt{3})$, this ratio is strictly bounded by
\begin{equation}
    \frac{\rho_{\text{corr}}}{\rho_{\text{bg}}} \leq \frac{32}{\sqrt{3}} \frac{c_6}{l^2} \, .
\end{equation}
Assuming the higher-derivative coupling is Planck-suppressed ($c_6 \sim m_P^{-2}$), this ratio scales proportionally to $(H/m_P)^2$. Given observational CMB constraints where the inflationary Hubble scale is $H \sim 10^{-5} m_P$, the geometric correction is subdominant by at least ten orders of magnitude ($\sim 10^{-10}$). Therefore, dropping the $r^{-3}$ term in the global slow-roll analysis is safely justified.

Overall, these results underscore that while the 3-form-driven setup can support phenomenologically viable slow-roll inflation, the available parameter space is tightly restricted. Notably, we find that thermodynamic consistency can impose constraints more stringent than those derived from inflationary dynamics alone. A successful inflationary trajectory in this model relies fundamentally on the precise calibration of the higher-derivative couplings to navigate the swampland-inspired thermodynamic bounds.


\section{Conclusion}

In this work, we explored the implications of thermodynamic positivity bounds in the context of a 3-form gauge field, with particular emphasis on both black hole physics and cosmological inflation. Starting from the classical analysis, we derived an extremality bound on the mass of black holes in a de Sitter background, showing that quantum gravity imposes non-trivial constraints on the allowable parameter space.

To further investigate the landscape/swampland separation, we introduced
higher-derivative corrections to the 3-form effective action and analyzed
them both geometrically and thermodynamically. Computing the correction to
the Wald entropy and imposing its positivity, $\Delta S > 0$, at the event
horizon yields a strict inequality on the higher-derivative couplings. We emphasize that this bound originates from
thermodynamic consistency alone and not from any decay or extremality
argument: evaluating the backreaction within the quasi-local thermodynamic
cavity bounded by the zero-force surface, we found that the correction to
the extremal mass vanishes identically, $\Delta M = 0$, so that the exact
Nariai state saturates the classical mass bound rather than being shifted
below it. The extremal limit is therefore where the bound is saturated,
not its origin; a non-zero shift appears only in the near-extremal regime,
where the cavity volume remains finite.
 
A central feature of this construction is that the resulting bound is
invariant under field redefinitions of the metric. Since the individual
couplings $c_i$ are not physical in a higher-derivative effective theory
--- a redefinition of the metric reshuffles them --- a constraint written
in terms of specific $c_i$ would carry no invariant meaning unless it can
be expressed through field-redefinition-invariant combinations. We showed that the entropy correction, and hence the
bound, is precisely of this form, with the physical content residing in
the invariant combination $c_{\text{inv}}$ together with $c_6$. This
invariance is, in our view, the decisive evidence that the constraint is a
genuine property of the effective theory rather than a re-expression of
the extremality condition on the mass in a particular field frame, and it
is what allows the same bound to be carried unchanged into the
cosmological setting. We found that the quasi-local boundary is essential to this
result. Terminating the backreaction integral at the zero-force surface is
what renders the entropy correction finite and expressible in invariant
combinations, so that the surface is not a mere regulator but the physical
boundary that isolates the localized black hole from the cosmological
background.


We then applied the same higher-order 3-form setup to study slow-roll inflation. In the large-field regime, the inflaton potential takes on a Higgs-like form, and we derived slow-roll conditions accordingly. In contrast, the small-field limit leads to a quartic effective potential with real extrema at a negative-energy (AdS) minimum, rendering it incompatible with the dS swampland criterion. Importantly, we demonstrated that thermodynamic consistency bounds could yield stricter constraints than those derived from conventional inflationary dynamics alone. This suggests that such thermodynamic criteria serve as powerful theoretical tools in guiding model building in the early universe.

Finally, it is important to acknowledge the limitations of the current framework. Our analysis heavily relies on the assumption of exact spherical symmetry and treats the underlying black hole geometry as a classical background upon which the higher-derivative effective field theory is formulated. A more complete quantum gravity treatment would require accounting for the quantum fluctuations of the metric itself, rather than treating them strictly as a backreaction from the 3-form gauge field. Furthermore, relaxing the assumption of spherical symmetry—for instance, by extending the analysis to rotating (Kerr-de Sitter) black holes—could introduce non-trivial couplings between the angular momentum and the higher-form field strength, potentially altering the thermodynamic positivity bounds. These caveats define the regime of validity for our present results and offer compelling avenues for future investigations.

A further caveat is dynamical rather than geometrical. Canonical single-field slow roll is
a regime we have chosen, not a property of the action: the operator $c_8$ already induces a
field-dependent kinetic function, and the canonical form is recovered only in the limits we
take. Two extensions suggest themselves. In warm inflation \cite{Berera:1995ie,Berera:2008ar},
dissipation into a radiation bath relaxes the slow-roll requirement and permits steeper
potentials, which has been argued to ease the de Sitter swampland tension discussed above
\cite{Motaharfar:2018zyb,Das:2018hqy}. We note that the thermodynamic bound derived here
would survive such a modification unchanged, since it constrains the couplings of the
effective action directly and is insensitive to the inflationary dynamics; only the
observationally viable contours of Figure~\ref{fig:c7vsg3} - \ref{fig:c6vsc} would move. Realising this
would nonetheless require supplementing the action with a coupling of the 3-form to light
matter, which our setup does not contain. Hyperinflation \cite{Brown:2017osf} requires at
least two fields on a negatively curved field space and is therefore inaccessible to a
single 3-form, although it has in any case been found to be in tension with the swampland
conditions rather than to relieve them \cite{Bjorkmo:2019qno}. We leave both directions to
future work.

Our results highlight the deep interplay between swampland-inspired consistency conditions, black hole thermodynamics, and inflationary cosmology. They suggest promising avenues for further exploration, particularly in refining the role of higher-form fields in connecting ultraviolet quantum gravity insights to observable phenomena.


\section*{Acknowledgement}
We are grateful to Daris Samart, Supakchai Ponglertsakul, Natthason Autthisin, and Jaturaporn Wattanakumpolkij for their insightful comments and stimulating discussions. This research project is supported by National Research Council of Thailand (NRCT) : (Contact No. N41A680320). CP is supported by Fundamental Fund 2568 of Khon Kaen University. This research has received funding support from the NSRF via the Program Management Unit for Human Resources \& Institutional Development, Research and Innovation [grant number B39G680009]
\appendix

\section{Energy-momentum tensor}
\label{A1}
The energy-momentum tensor is defined by the variational derivative of the matter action with respect to the metric: 
\begin{equation}
    T_{\mu\nu} = -\frac{2}{\sqrt{-g}}\frac{\delta S_{M}}{\delta g^{\mu\nu}},
\end{equation}
In our model, the matter action includes higher-order curvature and 3-form gauge field contributions. Accordingly, the full expression for the energy-momentum tensor involves corrections beyond the standard Einstein-Hilbert term. These higher-order contributions, denoted as 
\begin{equation}
    \begin{aligned}
        (\Delta T_{1})_{\mu}^{\ \nu} =& \delta_{\mu}^{\ \nu} R^2 - 4R R_{\mu}^{\ \nu} + 4 \nabla^\nu \nabla_\mu R - 4 \delta_{\mu}^{\ \nu}\Box R\\
         (\Delta T_{2})_{\mu}^{\ \nu} =& \delta_{\mu}^{\ \nu} R_{\rho \sigma} R^{\rho \sigma} + 4 \nabla_\alpha \nabla^\nu R^{\ \alpha}_{\mu} - 2 \Box R_{\mu}^{\ \nu} - \delta_{\mu}^{\ \nu} \Box R + 4 \nabla_\nu \nabla_\mu R \notag \\
       &+ 8 R^{\ \alpha}_\mu R_{\alpha}^{\ \nu} + 8 R^{\alpha \beta} R_{\alpha\beta\mu}^{\ \ \ \ \ \nu}\\
         (\Delta T_{3})_{\mu}^{\ \nu} =& \delta_{\mu}^{\ \nu} R_{\alpha \beta \gamma \delta} R^{\alpha \beta \gamma \delta} - 4 R_{\mu \alpha \beta \gamma} R^{\nu \alpha \beta \gamma}- 8 \Box R_{\mu}^{\ \nu}- 8 R^{\alpha \beta} R_{\mu \alpha\beta}^{\ \ \ \ \nu} \\
            (\Delta T_{4})_{\mu}^{\ \nu} =& -2F^2 R_{\mu}^{\ \nu}+\delta_{\mu}^{\ \nu}F^2 R- 8 F_{\mu}^{\ \sigma\rho\lambda}F^{\nu}_{\ \sigma\rho\lambda}-4\delta_{\mu}^{\ \nu}\nabla_{\lambda}(F^{\alpha\beta\gamma\delta}\nabla^{\lambda}F_{\alpha\beta\gamma\delta})\notag\\
       &+ 4 \nabla^{\nu}(F_{\alpha\beta\gamma\delta}\nabla_{\mu}F^{\alpha\beta\gamma\delta})\\
        (\Delta T_{5})_{\mu}^{\ \nu} =& -2F_{\sigma\rho\lambda\delta}(F^{\nu\rho\lambda\delta} R_{\mu}^{\ \sigma}+F_{\mu}^{\ \rho\lambda\delta}R^{\nu\sigma})+2R^{\sigma\rho}(\delta_{\mu}^{\ \nu}F_{\sigma}^{\ \lambda\delta\beta}F_{\rho\lambda\delta\beta}-3 F_{\mu\sigma}^{\ \ \lambda\delta}F^{\nu}_{\ \rho\lambda\delta})\\
        &-3 \delta_{\mu}^{\ \nu}F_{\sigma\lambda}^{\ \ \alpha\beta}F_{\rho\delta\alpha\beta}R^{\sigma\rho\lambda\delta}-\delta_{\mu}^{\ \nu}\nabla_{\lambda}\nabla^{\delta}(F^{\sigma\rho\beta\lambda}F_{\sigma\rho\beta\delta})\\
        &-\nabla_{\lambda}\big[\nabla_{\mu}(F^{\sigma\rho\beta\lambda}F^{\nu}_{\ \sigma\rho\beta}) + \nabla^{\nu}(F^{\sigma\rho\beta\lambda}F_{\mu\sigma\rho\beta})+\nabla^{\lambda}(F_{\mu\sigma\rho\beta}F^{\nu\sigma\rho\beta})        \big]\\
   \end{aligned}
\end{equation}

        \begin{equation}
    \begin{aligned}
         (\Delta T_{6})_{\mu}^{\ \nu} =& 4 F_{\mu\sigma}^{\ \ \rho\lambda}F^{\nu}_{\ \delta\rho\lambda}R^{\sigma\delta}+(2 F_{\rho}^{\ \nu\delta\beta}F_{\sigma\lambda\delta\beta}+3F_{\sigma}^{\ \nu\delta\beta}F_{\rho\lambda\delta\beta})R_{\mu}^{\ \sigma\rho\lambda}\\
         &- (2 F_{\mu\rho}^{\ \ \delta\beta}F_{\sigma\lambda\delta\beta}+3F_{\mu\lambda}^{\ \ \delta\beta}F_{\rho\lambda\delta\beta})R^{\nu\sigma\rho\lambda}-(8 F_{\mu\sigma\lambda}^{\ \ \ \beta}F^{\nu}_{\ \rho\delta\beta}+4F_{\mu\sigma\rho}^{\ \ \ \beta}F^{\nu}_{\ \lambda\delta\beta}\\
         &-\delta_{\mu}^{\ \nu}F_{\sigma\rho}^{\ \ \beta\gamma}F_{\lambda\delta\beta\gamma})R^{\sigma\rho\lambda\delta}-4\nabla_{\delta}(F^{\nu\sigma\rho\lambda}\nabla_{\lambda}F_{\mu\sigma\rho}^{\ \ \ \delta}+F_{\mu\sigma\rho}^{\ \ \ \lambda}\nabla_{\lambda}F^{\nu\sigma\rho\delta})\\
         &-4(\nabla_{\sigma}F_{\mu}^{\ \sigma\rho\lambda})(\nabla_{\delta}F^{\nu\delta}_{\ \ \rho\lambda})+4\nabla_{\delta}(F_{\mu}^{\ \sigma\rho\lambda})(\nabla_{\lambda}F^{\nu\delta}_{\ \ \sigma\rho})\\
         (\Delta T_{7})_{\mu}^{\ \nu} =& \delta_{\mu}^{\ \nu}F^4 - 16 F^2 F_{\mu\sigma\rho\delta}F^{\nu\sigma\rho\delta}\\
         (\Delta T_{8})_{\mu}^{\ \nu} =& \delta_{\mu}^{\ \nu}\nabla_{\beta}F_{\sigma\rho\lambda\delta}\nabla^{\beta}F^{\sigma\rho\lambda\delta}-2 \nabla_{\mu}F^{\sigma\rho\lambda\delta}\nabla^{\nu}F_{\sigma\rho\lambda\delta} - 4 \nabla_{\mu}F^{\nu\sigma\rho\lambda}\nabla_{\delta}F^{\delta}_{\ \sigma\rho\lambda}\\
         &+4 \nabla^{\nu}F_{\mu}^{\ \sigma\rho\lambda}\nabla_{\delta}F_{\sigma\rho\lambda}^{\ \  \ \delta} - 4 F^{\nu\sigma\rho\lambda}\nabla_{\delta}\nabla_{\mu}F_{\sigma\rho\lambda}^{\ \ \ \delta} + 4 F^{\sigma\rho\lambda\delta}\nabla_{\delta}\nabla^{\nu}F_{\mu\sigma\rho\delta}\\
         &- 4 F_{\mu}^{\ \sigma\rho\lambda}\nabla_{\delta}\nabla^{\nu}F_{\sigma\rho\lambda}^{\ \ \ \delta}+ 4 F^{\nu\sigma\rho\lambda}\Box F_{\mu\sigma\rho\lambda}+ 4 F_{\mu}^{\ \sigma\rho\lambda}\Box F^{\nu}_{\sigma\rho\lambda}\\
         &-4 \nabla^{\nu} F_{\sigma\rho\lambda\delta}\nabla^{\delta}F_{\mu}^{\ \sigma\rho\lambda}- 4 \nabla_{\mu}F_{\sigma\rho\lambda\delta}\nabla^{\delta}F^{\nu\sigma\rho\lambda},
    \end{aligned}
\end{equation}
where we defined $F^{2} = F_{\mu\nu\sigma\rho}F^{\mu\nu\sigma\rho}$ and $F^{4} = F_{\mu\nu\sigma\rho}F^{\mu\nu\sigma\rho} F_{\alpha\beta\gamma\delta}F^{\alpha\beta\gamma\delta}$.

\section{Detailed Derivations of Inflationary Potential and Slow-Roll Framework}
In this appendix, we provide the step-by-step mathematical derivations for the effective inflaton potential $V(\chi)$, the background equations of motion, and the slow-roll parameters presented in Section \ref{Cosmosec}.

\subsection{Equation of Motion for the Inflaton}
In the physical frame, the action for the canonical scalar field $\chi$ is given by Eq.\eqref{actionnew}. Varying this action with respect to $\chi$ yields the Euler-Lagrange equation:
\begin{equation}
\frac{1}{\sqrt{-\hat{g}}}\partial_\mu \left(\sqrt{-\hat{g}}\hat{g}^{\mu\nu}\partial_\nu \chi\right) - \frac{dV}{d\chi} = 0 \implies \hat{\Box}\chi - \frac{dV}{d\chi} = 0
\end{equation}
In a spatially homogeneous and isotropic Friedmann-Lema\^{i}tre-Robertson-Walker (FLRW) universe described by the metric $d\hat{s}^2 = -d\hat{t}^2 + \hat{a}(\hat{t})^2 \delta_{ij}dx^i dx^j$, the scalar field depends only on cosmic time ($\chi = \chi(\hat{t})$). The d'Alembertian operator reduces to $\hat{\Box}\chi = -\ddot{\chi} - 3H\dot{\chi}$, where $H \equiv \dot{\hat{a}}/\hat{a}$ is the Hubble parameter. Substituting this back into the field equation gives the standard cosmic-time evolution equation Eq.\eqref{eqX}:
\begin{equation}
\ddot{\chi} + 3H\dot{\chi} + \frac{dV}{d\chi} = 0
\end{equation}

\subsection{Standard Slow-Roll Parameters and e-folds}
Under the slow-roll approximation, the kinetic energy of the inflaton is assumed to be much smaller than its potential energy ($\frac{1}{2}\dot{\chi}^2 \ll V$), and the field acceleration is subdominant to the friction term ($\ddot{\chi} \ll 3H\dot{\chi}$). The equations of motion reduce to:
\begin{equation}
3H\dot{\chi} \approx -\frac{dV}{d\chi}, \quad 3H^2 \approx \frac{V}{m_P^2}
\label{app:slow_roll_bg}
\end{equation}
The conditions for the validity of this approximation are parameterized by the potential slow-roll parameters $\epsilon_\chi$ and $\eta_\chi$, defined geometrically by the slope and curvature of the potential, leading to Eqs. \eqref{epsix} and \eqref{etax}:
\begin{equation}
\epsilon_\chi \equiv \frac{m_P^2}{2}\left(\frac{V'(\chi)}{V(\chi)}\right)^2, \quad \eta_\chi \equiv m_P^2\left(\frac{V''(\chi)}{V(\chi)}\right)
\end{equation}
The total number of e-folds $N_\chi$ from the initial field value $\chi_i$ to the end of inflation $\chi_e$ is defined by the integral of the scale factor growth $N_\chi = \int_{\hat{t}_i}^{\hat{t}_e} H d\hat{t}$. Transforming the integration variable from $\hat{t}$ to $\chi$ via $d\hat{t} = \frac{d\chi}{\dot{\chi}}$, and using the slow-roll relation $\dot{\chi} \approx -V'/(3H)$ from Eq. (\ref{app:slow_roll_bg}), we find:
\begin{equation}
N_\chi = \int_{\chi_i}^{\chi_e} H \left(-\frac{3H}{V'}\right) d\chi = -\int_{\chi_i}^{\chi_e} \frac{3H^2}{V'} d\chi
\end{equation}
Substituting $3H^2 \approx V/m_P^2$ into the integrand and reversing the limits of integration to maintain a positive e-fold count yields Eq. \eqref{efold}:
\begin{equation}
\label{eq:efolds_proof}
N_\chi = \frac{1}{m_P^2} \int_{\chi_e}^{\chi_i} \frac{V(\chi)}{V'(\chi)} d\chi
\end{equation}

\subsection{Derivation of the Tensor-to-Scalar Ratio}
\label{app:tensor_scalar_ratio}
In this appendix, we briefly review the standard derivation of the tensor-to-scalar ratio in canonical single-field slow-roll inflation. The action is given by
\begin{equation}
S = \int d^4x \sqrt{-g}
\left[
\frac{m_{ P}^2}{2}R
-\frac{1}{2}\partial_{\mu}\chi\partial^{\mu}\chi
-V(\chi)
\right],
\end{equation}
where \(m_P\) is the reduced Planck mass and \(\chi\) denotes the inflaton field. The first slow-roll parameter is defined as
\begin{equation}
\epsilon \equiv -\frac{\dot H}{H^2}.
\end{equation}
During slow-roll inflation, one has \(\epsilon \ll 1\), together with the approximate background equations
\begin{equation}
3m_{P}^2H^2 \simeq V,
\qquad
3H\dot{\chi} \simeq -V_{,\chi}.
\end{equation}
The corresponding potential slow-roll parameter is
\begin{equation}
\epsilon_V
=
\frac{m_P^2}{2}
\left(
\frac{V_{,\chi}}{V}
\right)^2,
\end{equation}
which satisfies \(\epsilon \simeq \epsilon_V\) in the slow-roll regime.
The scalar curvature perturbation \(\mathcal R\) is described, at quadratic order, by
\begin{equation}
S_{\mathcal R}^{(2)}
=
m_P^2
\int d\tau d^3x\,
a^2\epsilon
\left[
(\mathcal R')^2
-
(\nabla \mathcal R)^2
\right],
\end{equation}
where \(\tau\) is conformal time and a prime denotes differentiation with respect to \(\tau\). Quantizing the canonically normalized scalar mode gives the scalar power spectrum at horizon crossing \(k=aH\),
\begin{equation}
 P_{s}
=
\frac{H^2}{8\pi^2m_P^2\epsilon}.
\label{eq:scalar_power_spectrum}
\end{equation}
Tensor perturbations are introduced through the spatial metric as
\begin{equation}
ds^2
=
-dt^2
+
a^2(t)
\left(
\delta_{ij}+h_{ij}
\right)dx^idx^j,
\end{equation}
where \(h_{ij}\) is transverse and traceless,
\begin{equation}
\partial_i h_{ij}=0,
\qquad
h_{ii}=0.
\end{equation}
The tensor perturbation contains two physical polarizations. The corresponding tensor power spectrum is
\begin{equation}
P_T
=
\frac{2H^2}{\pi^2m_P^2}.
\label{eq:tensor_power_spectrum}
\end{equation}
The factor of \(2\) accounts for the two tensor polarizations.
The tensor-to-scalar ratio is defined by
\begin{equation}
r \equiv \frac{P_T}{P_{s}}.
\end{equation}
Using Eqs.~\eqref{eq:scalar_power_spectrum} and \eqref{eq:tensor_power_spectrum}, one obtains
\begin{equation}
r
=
\frac{
\dfrac{2H^2}{\pi^2m_P^2}
}{
\dfrac{H^2}{8\pi^2m_P^2\epsilon}
}
=
16\epsilon.
\end{equation}
Thus, for canonical single-field slow-roll inflation,
\begin{equation}
r = 16\epsilon
\end{equation}
Equivalently, in terms of the potential slow-roll parameter,
\begin{equation}
r \simeq 16\epsilon_V
=
8m_P^2
\left(
\frac{V_{,\chi}}{V}
\right)^2.
\end{equation}

\subsection{Scalar Spectral Index}
The primordial scalar power spectrum $P_s(k)$ can be expanded as a power law around a pivot scale $k_*$. The scalar spectral index $n_s$ is defined by the scale dependence of the power spectrum:
\begin{equation}
n_s - 1 \equiv \frac{d \ln P_s}{d \ln k}
\end{equation}
Using the relation $d\ln k \approx dN_\chi = \frac{1}{m_P^2}\frac{V}{V'}d\chi$ at horizon exit ($k=aH$), and expressing the amplitude of the scalar perturbations in the slow-roll limit as $P_s \propto \frac{V^3}{m_P^6 (V')^2}$, we evaluate the derivative using the chain rule:
\begin{equation}
n_s - 1 = \frac{d \ln P_s}{d\chi} \frac{d\chi}{d\ln k} = m_P^2 \frac{V'}{V} \frac{d}{d\chi} \left[ \ln\left(\frac{V^3}{(V')^2}\right) \right]
\end{equation}
Differentiating the logarithmic term gives:
\begin{equation}
\frac{d}{d\chi} \left( 3\ln V - 2\ln V' \right) = \frac{3V'}{V} - \frac{2V''}{V'}
\end{equation}
Multiplying through by $m_P^2 \frac{V'}{V}$, we obtain:
\begin{equation}
n_s - 1 = m_P^2 \left[ 3\left(\frac{V'}{V}\right)^2 - 2\frac{V''}{V} \right]
\end{equation}
Recognizing that $3\left(\frac{V'}{V}\right)^2 = 6\left[\frac{m_P^2}{2}\left(\frac{V'}{V}\right)^2\right]\frac{1}{m_P^2} = \frac{6\epsilon_\chi}{m_P^2}$ and $\frac{V''}{V} = \frac{\eta_\chi}{m_P^2}$, we substitute the definitions of the slow-roll parameters to arrive at Eq. (4.24):
\begin{equation}
n_s = 1 - 6\epsilon_\chi + 2\eta_\chi
\end{equation}



\bibliographystyle{JHEP}
\bibliography{biblio.bib}

@article{Barros:2020ghz,
    author = "Barros, Bruno J. and Dǎnilǎ, Bogdan and Harko, Tiberiu and Lobo, Francisco S. N.",
    title = "{Black hole and naked singularity geometries supported by three-form fields}",
    eprint = "2004.06605",
    archivePrefix = "arXiv",
    primaryClass = "gr-qc",
    doi = "10.1140/epjc/s10052-020-8178-1",
    journal = "Eur. Phys. J. C",
    volume = "80",
    pages = "617",
    year = "2020"
}

@article{Fernando:2013mex,
    author = "Fernando, Sharmanthie",
    title = "{Nariai black holes with quintessence}",
    eprint = "1408.5064",
    archivePrefix = "arXiv",
    primaryClass = "gr-qc",
    doi = "10.1142/S0217732313501897",
    journal = "Mod. Phys. Lett. A",
    volume = "28",
    pages = "1350189",
    year = "2013"
}

@article{LOUSTO1988411,
title = {Back reaction effects in black hole spacetimes},
journal = {Physics Letters B},
volume = {212},
number = {4},
pages = {411-414},
year = {1988},
issn = {0370-2693},
doi = {https://doi.org/10.1016/0370-2693(88)91789-3},
url = {https://www.sciencedirect.com/science/article/pii/0370269388917893},
author = {C.O. Loustó and N. Sánchez}
}

@article{Mishra:2018dtg,
    author = "Mishra, Swagat S. and Sahni, Varun and Toporensky, Alexey V.",
    title = "{Initial conditions for Inflation in an FRW Universe}",
    eprint = "1801.04948",
    archivePrefix = "arXiv",
    primaryClass = "gr-qc",
    doi = "10.1103/PhysRevD.98.083538",
    journal = "Phys. Rev. D",
    volume = "98",
    number = "8",
    pages = "083538",
    year = "2018"
}

@article{Vafa:2005ui,
    author = "Vafa, Cumrun",
    title = "{The String landscape and the swampland}",
    eprint = "hep-th/0509212",
    archivePrefix = "arXiv",
    reportNumber = "HUTP-05-A043",
    month = "9",
    year = "2005"
}

@article{Arkani-Hamed:2006emk,
    author = "Arkani-Hamed, Nima and Motl, Lubos and Nicolis, Alberto and Vafa, Cumrun",
    title = "{The String landscape, black holes and gravity as the weakest force}",
    eprint = "hep-th/0601001",
    archivePrefix = "arXiv",
    reportNumber = "HUTP-05-A0057",
    doi = "10.1088/1126-6708/2007/06/060",
    journal = "JHEP",
    volume = "06",
    pages = "060",
    year = "2007"
}

@article{Kats:2006xp,
    author = "Kats, Yevgeny and Motl, Lubos and Padi, Megha",
    title = "{Higher-order corrections to mass-charge relation of extremal black holes}",
    eprint = "hep-th/0606100",
    archivePrefix = "arXiv",
    reportNumber = "HUTP-06-A0023",
    doi = "10.1088/1126-6708/2007/12/068",
    journal = "JHEP",
    volume = "12",
    pages = "068",
    year = "2007"
}

@article{Cheung:2018cwt,
    author = "Cheung, Clifford and Liu, Junyu and Remmen, Grant N.",
    title = "{Proof of the Weak Gravity Conjecture from Black Hole Entropy}",
    eprint = "1801.08546",
    archivePrefix = "arXiv",
    primaryClass = "hep-th",
    reportNumber = "CALT-TH-2018-007",
    doi = "10.1007/JHEP10(2018)004",
    journal = "JHEP",
    volume = "10",
    pages = "004",
    year = "2018"
}

@article{Ovrut:1997ur,
    author = "Ovrut, Burt A. and Waldram, Daniel",
    title = "{Membranes and three form supergravity}",
    eprint = "hep-th/9704045",
    archivePrefix = "arXiv",
    reportNumber = "UPR-0741-T, IASSNS-HEP-97-26, PUPT-1691",
    doi = "10.1016/S0550-3213(97)00510-5",
    journal = "Nucl. Phys. B",
    volume = "506",
    pages = "236--266",
    year = "1997"
}

@article{Bandos:2018gjp,
    author = "Bandos, Igor and Farakos, Fotis and Lanza, Stefano and Martucci, Luca and Sorokin, Dmitri",
    title = "{Three-forms, dualities and membranes in four-dimensional supergravity}",
    eprint = "1803.01405",
    archivePrefix = "arXiv",
    primaryClass = "hep-th",
    reportNumber = "DFPD-2018/TH/01",
    doi = "10.1007/JHEP07(2018)028",
    journal = "JHEP",
    volume = "07",
    pages = "028",
    year = "2018"
}

@article{Koivisto:2009fb,
    author = "Koivisto, Tomi S. and Nunes, Nelson J.",
    title = "{Inflation and dark energy from three-forms}",
    eprint = "0908.0920",
    archivePrefix = "arXiv",
    primaryClass = "astro-ph.CO",
    doi = "10.1103/PhysRevD.80.103509",
    journal = "Phys. Rev. D",
    volume = "80",
    pages = "103509",
    year = "2009"
}

@article{Germani:2009iq,
    author = "Germani, Cristiano and Kehagias, Alex",
    title = "{P-nflation: generating cosmic Inflation with p-forms}",
    eprint = "0902.3667",
    archivePrefix = "arXiv",
    primaryClass = "astro-ph.CO",
    doi = "10.1088/1475-7516/2009/03/028",
    journal = "JCAP",
    volume = "03",
    pages = "028",
    year = "2009"
}

@article{Mulryne:2012ax,
    author = "Mulryne, David J. and Noller, Johannes and Nunes, Nelson J.",
    title = "{Three-form inflation and non-Gaussianity}",
    eprint = "1209.2156",
    archivePrefix = "arXiv",
    primaryClass = "astro-ph.CO",
    doi = "10.1088/1475-7516/2012/12/016",
    journal = "JCAP",
    volume = "12",
    pages = "016",
    year = "2012"
}

@article{Kaloper:2011jz,
    author = "Kaloper, Nemanja and Lawrence, Albion and Sorbo, Lorenzo",
    title = "{An Ignoble Approach to Large Field Inflation}",
    eprint = "1101.0026",
    archivePrefix = "arXiv",
    primaryClass = "hep-th",
    doi = "10.1088/1475-7516/2011/03/023",
    journal = "JCAP",
    volume = "03",
    pages = "023",
    year = "2011"
}

@article{Wald:1993nt,
    author = "Wald, Robert M.",
    title = "{Black hole entropy is the Noether charge}",
    eprint = "gr-qc/9307038",
    archivePrefix = "arXiv",
    reportNumber = "EFI-93-42",
    doi = "10.1103/PhysRevD.48.R3427",
    journal = "Phys. Rev. D",
    volume = "48",
    number = "8",
    pages = "R3427--R3431",
    year = "1993"
}

@article{Bodendorfer:2013wga,
    author = "Bodendorfer, Norbert and Neiman, Yasha",
    title = "{Wald entropy formula and loop quantum gravity}",
    eprint = "1304.3025",
    archivePrefix = "arXiv",
    primaryClass = "gr-qc",
    reportNumber = "IGC-13-4-2",
    doi = "10.1103/PhysRevD.90.084054",
    journal = "Phys. Rev. D",
    volume = "90",
    number = "8",
    pages = "084054",
    year = "2014"
}

@article{Planck:2018jri,
    author = "Akrami, Y. and others",
    collaboration = "Planck",
    title = "{Planck 2018 results. X. Constraints on inflation}",
    eprint = "1807.06211",
    archivePrefix = "arXiv",
    primaryClass = "astro-ph.CO",
    doi = "10.1051/0004-6361/201833887",
    journal = "Astron. Astrophys.",
    volume = "641",
    pages = "A10",
    year = "2020"
}

@article{Antoniadis:2020xso,
    author = "Antoniadis, Ignatios and Benakli, Karim",
    title = "{Weak Gravity Conjecture in de Sitter Space-Time}",
    eprint = "2006.12512",
    archivePrefix = "arXiv",
    primaryClass = "hep-th",
    doi = "10.1002/prop.202000054",
    journal = "Fortsch. Phys.",
    volume = "68",
    number = "9",
    pages = "2000054",
    year = "2020"
}

@article{Obied:2018sgi,
    author = "Obied, Georges and Ooguri, Hirosi and Spodyneiko, Lev and Vafa, Cumrun",
    title = "{De Sitter Space and the Swampland}",
    eprint = "1806.08362",
    archivePrefix = "arXiv",
    primaryClass = "hep-th",
    reportNumber = "CALT-TH-2018-020, IPMU18-0100",
    month = "6",
    year = "2018"
}

@article{Agrawal:2018own,
    author = "Agrawal, Prateek and Obied, Georges and Steinhardt, Paul J. and Vafa, Cumrun",
    title = "{On the Cosmological Implications of the String Swampland}",
    eprint = "1806.09718",
    archivePrefix = "arXiv",
    primaryClass = "hep-th",
    doi = "10.1016/j.physletb.2018.07.040",
    journal = "Phys. Lett. B",
    volume = "784",
    pages = "271--276",
    year = "2018"
}

@article{Ooguri:2018wrx,
    author = "Ooguri, Hirosi and Palti, Eran and Shiu, Gary and Vafa, Cumrun",
    title = "{Distance and de Sitter Conjectures on the Swampland}",
    eprint = "1810.05506",
    archivePrefix = "arXiv",
    primaryClass = "hep-th",
    doi = "10.1016/j.physletb.2018.11.018",
    journal = "Phys. Lett. B",
    volume = "788",
    pages = "180--184",
    year = "2019"
}

@article{Chang-Young:2010lou,
    author = "Chang-Young, Ee and Eune, Myungseok and Kimm, Kyoungtae and Lee, Daeho",
    title = "{Schwarzschild-de Sitter black hole from entropic viewpoint}",
    eprint = "1011.3960",
    archivePrefix = "arXiv",
    primaryClass = "hep-th",
    doi = "10.1142/S0217732311036450",
    journal = "Mod. Phys. Lett. A",
    volume = "26",
    pages = "1975--1983",
    year = "2011"
}

@article{Cremonini:2019xue,
  author        = {Cremonini, Sera and Jones, Callum R. T. and Liu, James T. and McPeak, Brian},
  title         = {{Higher-Derivative Corrections to Entropy and the Weak Gravity Conjecture in Anti-de Sitter Space}},
  eprint        = {1912.11161},
  archivePrefix = {arXiv},
  primaryClass  = {hep-th},
  doi           = {10.1007/JHEP09(2020)003},
  journal       = {JHEP},
  volume        = {09},
  pages         = {003},
  year          = {2020}
}

@article{Goon:2019faz,
  author        = {Goon, Garrett and Penco, Riccardo},
  title         = {{Universal Relation between Corrections to Entropy and Extremality}},
  eprint        = {1909.05254},
  archivePrefix = {arXiv},
  primaryClass  = {hep-th},
  doi           = {10.1103/PhysRevLett.124.101103},
  journal       = {Phys. Rev. Lett.},
  volume        = {124},
  pages         = {101103},
  year          = {2020}
}

@article{McPeak:2021gpt,
  author        = {McPeak, Brian},
  title         = {{Higher-derivative corrections to black hole entropy at zero temperature}},
  eprint        = {2112.13433},
  archivePrefix = {arXiv},
  primaryClass  = {hep-th},
  doi           = {10.1103/PhysRevD.105.L081901},
  journal       = {Phys. Rev. D},
  volume        = {105},
  number        = {8},
  pages         = {L081901},
  year          = {2022}
}

@article{Reall:2019sec,
  author        = {Reall, Harvey S. and Santos, Jorge E.},
  title         = {{Higher derivative corrections to Kerr black hole thermodynamics}},
  eprint        = {1901.11535},
  archivePrefix = {arXiv},
  primaryClass  = {hep-th},
  doi           = {10.1007/JHEP04(2019)021},
  journal       = {JHEP},
  volume        = {04},
  pages         = {021},
  year          = {2019}
}

@article{Hamada:2018dde,
  author        = {Hamada, Yuta and Noumi, Toshifumi and Shiu, Gary},
  title         = {{Weak Gravity Conjecture from Unitarity and Causality}},
  eprint        = {1810.03637},
  archivePrefix = {arXiv},
  primaryClass  = {hep-th},
  doi           = {10.1103/PhysRevLett.123.051601},
  journal       = {Phys. Rev. Lett.},
  volume        = {123},
  pages         = {051601},
  year          = {2019}
}

@article{Loges:2020trf,
  author        = {Loges, Gabriel J. and Noumi, Toshifumi and Shiu, Gary},
  title         = {{Thermodynamics of 4D Dilatonic Black Holes and the Weak Gravity Conjecture}},
  eprint        = {1909.01352},
  archivePrefix = {arXiv},
  primaryClass  = {hep-th},
  doi           = {10.1103/PhysRevD.102.046010},
  journal       = {Phys. Rev. D},
  volume        = {102},
  pages         = {046010},
  year          = {2020}
}

@article{Cheung:2016wjt,
  author        = {Cheung, Clifford and Remmen, Grant N.},
  title         = {{Positivity of Curvature-Squared Corrections in Gravity}},
  eprint        = {1608.02942},
  archivePrefix = {arXiv},
  primaryClass  = {hep-th},
  doi           = {10.1103/PhysRevLett.118.051601},
  journal       = {Phys. Rev. Lett.},
  volume        = {118},
  pages         = {051601},
  year          = {2017}
}

@article{Barros:2015evi,
  author        = {Barros, Bruno J. and Nunes, Nelson J.},
  title         = {{Three-form inflation in type II Randall-Sundrum}},
  eprint        = {1511.07856},
  archivePrefix = {arXiv},
  primaryClass  = {astro-ph.CO},
  doi           = {10.1103/PhysRevD.93.043512},
  journal       = {Phys. Rev. D},
  volume        = {93},
  pages         = {043512},
  year          = {2016}
}

@article{NooriGashti:2025modmax,
  author        = {Noori Gashti, Saeed and Afshar, Mohammad Ali S. and Alipour, Mohammad Reza and Sakall{\i}, {\.I}zzet and Pourhassan, Behnam and Sadeghi, Jafar},
  title         = {{Weak gravity conjecture in ModMax black holes: weak cosmic censorship and photon sphere analysis}},
  eprint        = {2504.11939},
  archivePrefix = {arXiv},
  primaryClass  = {gr-qc},
  doi           = {10.1140/epjc/s10052-025-14890-8},
  journal       = {Eur. Phys. J. C},
  volume        = {85},
  pages         = {1144},
  year          = {2025}
}

@article{NooriGashti:2024rastall,
  author        = {Noori Gashti, Saeed and Sakall{\i}, {\.I}zzet and Pourhassan, Behnam},
  title         = {{Thermodynamic topology, photon spheres, and evidence for weak gravity conjecture in charged black holes with perfect fluid within Rastall theory}},
  eprint        = {2410.14492},
  archivePrefix = {arXiv},
  primaryClass  = {gr-qc},
  doi           = {10.1016/j.physletb.2025.139862},
  journal       = {Phys. Lett. B},
  volume        = {869},
  pages         = {139862},
  year          = {2025}
}

@article{Sadeghi:2023dstring,
  author        = {Sadeghi, Jafar and Pourhassan, Behnam and Noori Gashti, Saeed and Sakall{\i}, {\.I}zzet and Alipour, Mohammad Reza},
  title         = {{de Sitter swampland conjecture in string field inflation}},
  eprint        = {2303.04551},
  archivePrefix = {arXiv},
  primaryClass  = {gr-qc},
  doi           = {10.1140/epjc/s10052-023-11774-2},
  journal       = {Eur. Phys. J. C},
  volume        = {83},
  pages         = {635},
  year          = {2023}
}

@article{Sadeghi:2024wgc,
  author        = {Sadeghi, Jafar and Noori Gashti, Saeed and Sakall{\i}, {\.I}zzet and Pourhassan, Behnam},
  title         = {{Weak gravity conjecture of charged-rotating-AdS black hole surrounded by quintessence and string cloud}},
  eprint        = {2011.05109},
  archivePrefix = {arXiv},
  primaryClass  = {gr-qc},
  doi           = {10.1016/j.nuclphysb.2024.116581},
  journal       = {Nucl. Phys. B},
  volume        = {1004},
  pages         = {116581},
  year          = {2024}
}

@article{Berera:1995ie,
    author  = "Berera, Arjun",
    title   = "{Warm inflation}",
    journal = "Phys. Rev. Lett.",
    volume  = "75",
    pages   = "3218--3221",
    year    = "1995",
    eprint  = "astro-ph/9509049",
    archivePrefix = "arXiv",
    doi     = "10.1103/PhysRevLett.75.3218"
}

@article{Berera:2008ar,
    author  = "Berera, Arjun and Moss, Ian G. and Ramos, Rudnei O.",
    title   = "{Warm Inflation and its Microphysical Basis}",
    journal = "Rept. Prog. Phys.",
    volume  = "72",
    pages   = "026901",
    year    = "2009",
    eprint  = "0808.1855",
    archivePrefix = "arXiv",
    primaryClass  = "hep-ph",
    doi     = "10.1088/0034-4885/72/2/026901"
}

@article{Motaharfar:2018zyb,
    author  = "Motaharfar, Meysam and Kamali, Vahid and Ramos, Rudnei O.",
    title   = "{Warm inflation as a way out of the swampland}",
    journal = "Phys. Rev. D",
    volume  = "99",
    number  = "6",
    pages   = "063513",
    year    = "2019",
    eprint  = "1810.02816",
    archivePrefix = "arXiv",
    primaryClass  = "astro-ph.CO",
    doi     = "10.1103/PhysRevD.99.063513"
}

@article{Das:2018hqy,
    author  = "Das, Suratna",
    title   = "{Warm Inflation in the light of Swampland Criteria}",
    journal = "Phys. Rev. D",
    volume  = "99",
    number  = "6",
    pages   = "063514",
    year    = "2019",
    eprint  = "1810.05038",
    archivePrefix = "arXiv",
    primaryClass  = "hep-th",
    doi     = "10.1103/PhysRevD.99.063514"
}

@article{Brown:2017osf,
    author  = "Brown, Adam R.",
    title   = "{Hyperbolic Inflation}",
    journal = "Phys. Rev. Lett.",
    volume  = "121",
    number  = "25",
    pages   = "251601",
    year    = "2018",
    eprint  = "1705.03023",
    archivePrefix = "arXiv",
    primaryClass  = "hep-th",
    doi     = "10.1103/PhysRevLett.121.251601"
}

@article{Bjorkmo:2019qno,
    author  = "Bjorkmo, Theodor and Marsh, M. C. David",
    title   = "{Hyperinflation generalised: from its attractor mechanism to its tension with the `swampland conditions'}",
    journal = "JHEP",
    volume  = "04",
    pages   = "172",
    year    = "2019",
    eprint  = "1901.08603",
    archivePrefix = "arXiv",
    primaryClass  = "hep-th",
    doi     = "10.1007/JHEP04(2019)172"
}

@article{Aurilia:1980xj,
    author  = "Aurilia, A. and Nicolai, H. and Townsend, P. K.",
    title   = "{Hidden Constants: The Theta Parameter of QCD and the Cosmological Constant of N=8 Supergravity}",
    journal = "Nucl. Phys. B",
    volume  = "176",
    pages   = "509--522",
    year    = "1980",
    doi     = "10.1016/0550-3213(80)90466-6"
}

@article{Dvali:2005an,
    author  = "Dvali, Gia",
    title   = "{Three-form gauging of axion symmetries and gravity}",
    eprint  = "hep-th/0507215",
    archivePrefix = "arXiv",
    year    = "2005"
}

@article{Kaloper:2008fb,
    author  = "Kaloper, Nemanja and Sorbo, Lorenzo",
    title   = "{A Natural Framework for Chaotic Inflation}",
    journal = "Phys. Rev. Lett.",
    volume  = "102",
    pages   = "121301",
    year    = "2009",
    eprint  = "0811.1989",
    archivePrefix = "arXiv",
    primaryClass  = "hep-th",
    doi     = "10.1103/PhysRevLett.102.121301"
}

\end{document}